\documentclass[aps,prd,nofootinbib,superscriptaddress,onecolumn,notitlepage,preprintnumbers]{revtex4-1}
\pdfoutput=1

\usepackage{amsmath,graphicx,verbatim,epsfig}
\usepackage{epstopdf}
\usepackage[utf8x]{inputenc}
\usepackage{amsmath,graphicx}
\usepackage{simplewick}
\usepackage{datetime}
\usepackage[colorlinks=true,linkcolor=blue,citecolor=blue]{hyperref}
\usepackage[usenames,dvipsnames]{color}
\usepackage{setspace}

\newcommand{\eps}{\varepsilon}

\newcommand{\Ga}{\Gamma}

\newcommand{\nn}{\nonumber}

\newcommand{\bn}{{\bar n}}

\newcommand{\pslash}{{\not \!p}}
\newcommand{\kslash}{{\not \!k}}

\newcommand{\bnslash}{{\not \!\bn}}

\newcommand{\nb}{\bar n}

\newcommand{\veuv}{\varepsilon_{\rm {UV}}}

\newcommand{\be}{\begin{equation}}
\newcommand{\ee}{\end{equation}}
\newcommand{\bea}{\begin{eqnarray}}
\newcommand{\eea}{\end{eqnarray}}
\newcommand{\balign}{\begin{align}}
\newcommand{\ealign}{\end{align}}
\newcommand{\as}{\alpha_s}

\newcommand{\cd}{\cdot}

\newcommand{\mB}{\mathcal{B}}

\newcommand{\ket}[1]{\left | #1 \right >}

\newcommand{\sandwich}[3]{\left< #1 \right | #2 \left | #3 \right >}

\newcommand{\bg}{\begin{gather}}
\newcommand{\foma}{\end{gather}}

\newcommand{\noopsort}[1]{}

\newcommand{\vecb}[1]{\mbox{\boldmath $#1$}}
\newcommand{\vecbe}[1]{\mbox{\boldmath ${\scriptstyle #1}$}}
\newcommand{\vecbp}[1]{\mbox{\boldmath $#1_\perp$}}

\def\e{\epsilon}

\def\ve{\varepsilon}

\def\L{\Lambda}

\def\z{\zeta}

\def\<{\langle}
\def\>{\rangle}

\def\a{\alpha}
\def\b{\beta}
\def\g{\gamma}  \def\G{\Ga}
\def\d{\delta}  \def\D{\Delta}
\def\l{\lambda}   \def\L{\Lambda}
\def\s{\sigma}
\def\r{\rho}  
\def\x{\xi}

\def\m{\mu}
\def\n{\nu}

\def\z{\zeta}

\def\({\left(}
\def\[{\left[}
\def\){\right)}
\def\]{\right]}

\def\ln{\hbox{ln}}

\def\inf{\infty}

\def\bnslash{\bar n\!\!\!\slash}
\def\pslash{p\!\!\!\slash}

\def\kslash{k\!\!\!\slash}

\def\le{\left }
\def\ri{\right}
\def\bp{\bar p}
\def\bP{\bar P}

\def\gev{\rm GeV}

\def\lqcd{\L_{\rm QCD}}

\newcommand{\ben}{\begin{eqnarray}}
\newcommand{\een}{\end{eqnarray}}

\newcommand{\bef}{\begin{figure}[htb]\centering}
\newcommand{\eef}{\end{figure}}

\newcommand{\eq}[1]{Eq.~\eqref{#1}}

\begin{document}

\title{
QCD evolution of (un)polarized gluon TMDPDFs and the Higgs $q_T$-distribution}

\author{Miguel G. Echevarria}
\email{m.g.echevarria@nikhef.nl}
\affiliation{Nikhef Theory Group,
Science Park 105, 1098XG Amsterdam, the Netherlands}
\affiliation{Department of Physics and Astronomy, 
VU University Amsterdam, 
De Boelelaan 1081, NL-1081 HV Amsterdam, the Netherlands}
\author{Tomas Kasemets}
\email{kasemets@nikhef.nl}
\affiliation{Nikhef Theory Group,
Science Park 105, 1098XG Amsterdam, the Netherlands}
\affiliation{Department of Physics and Astronomy, 
VU University Amsterdam, 
De Boelelaan 1081, NL-1081 HV Amsterdam, the Netherlands}
\author{Piet J. Mulders}
\email{mulders@few.vu.nl}
\affiliation{Nikhef Theory Group,
Science Park 105, 1098XG Amsterdam, the Netherlands}
\affiliation{Department of Physics and Astronomy, 
VU University Amsterdam, 
De Boelelaan 1081, NL-1081 HV Amsterdam, the Netherlands}
\author{Cristian Pisano}
\email{c.pisano@nikhef.nl}
\affiliation{Nikhef Theory Group,
Science Park 105, 1098XG Amsterdam, the Netherlands}
\affiliation{Department of Physics and Astronomy, 
VU University Amsterdam, 
De Boelelaan 1081, NL-1081 HV Amsterdam, the Netherlands}
\affiliation{Department of Physics, University of Antwerp, 
Groenenborgerlaan 171, 2020 Antwerp, Belgium}




\begin{abstract}

We provide the proper definition of all the leading-twist (un)polarized gluon transverse momentum dependent parton distribution functions (TMDPDFs), by considering the Higgs boson transverse momentum distribution in hadron-hadron collisions and deriving the factorization theorem in terms of them.
We show that the evolution of all the (un)polarized gluon TMDPDFs is driven by a universal evolution kernel, which can be resummed up to next-to-next-to-leading-logarithmic accuracy.
Considering the proper definition of gluon TMDPDFs, we perform an explicit next-to-leading-order calculation of the unpolarized ($f_1^g$), linearly polarized ($h_1^{\perp g}$) and  helicity ($g_{1L}^g$) gluon TMDPDFs, and show that, as expected, they are free from rapidity divergences.
As a byproduct, we obtain the Wilson coefficients of the refactorization of these TMDPDFs at large transverse momentum.
In particular, the coefficient of $g_{1L}^g$, which has never been calculated before, constitutes a new and necessary ingredient for a reliable phenomenological extraction of this quantity, for instance at RHIC or the future AFTER@LHC or Electron-Ion Collider.
The coefficients of $f_1^g$ and $h_1^{\perp g}$ have never been calculated in the present formalism, although they could be obtained by carefully collecting and recasting previous results in the new TMD formalism.
We apply these results to analyze the contribution of linearly polarized gluons at different scales, relevant, for instance, for the inclusive production of the Higgs boson and the $C$-even pseudoscalar bottomonium state $\eta_{b}$.
Applying our resummation scheme we finally provide predictions for the Higgs boson $q_T$-distribution at the LHC.

\end{abstract}

\preprint{NIKHEF 2014-036}
\maketitle


\section{Introduction}
\label{sec:intro}
Observables sensitive to the transverse momentum of quarks and gluons inside a hadron have a long theoretical and experimental history. They have proven to be valuable tools to test the QCD dynamics at high-energy colliders, extending the information provided by observables integrated over the intrinsic transverse momenta.
At large transverse momentum these observables can be computed in perturbation theory, but if the transverse momentum $q_T$ is much smaller than the probe of the hard reaction $Q$, then large logarithms of their ratio appear and resummation becomes a must in order to obtain reliable results.
This issue was already addressed in the eighties by Collins, Soper and Sterman~\cite{Collins:1984kg}.

The main hadronic quantities in observables at $q_T\ll Q$ are the transverse momentum dependent functions (TMDs), first considered by Ralston and Soper~\cite{Ralston:1979ys,Ralston:1980pp} and by Collins and Soper~\cite{Collins:1981uw,Collins:1981uk}. The TMDs represent, generally speaking, the probability of finding a parton inside a hadron with a definite transverse momentum, i.e., TMD parton distribution functions (TMDPDFs); or the probability that a quark or gluon fragments into a hadron with a given transverse momentum (TMDFFs). They play an important role in the rich phenomenology of azimuthal and spin asymmetries (see, e.g., ~\cite{Mulders:1995dh,Bacchetta:2006tn}).

After the pioneering works, much effort has been devoted to properly describe the polarization of the partons/hadrons, the universality of TMDs and other relevant properties.
However, the ``naive'' (old) definitions introduced in~\cite{Collins:1981uw,Collins:1981uk} and considered in subsequent works, suffer from undesired features preventing them from properly represent physical hadronic quantities, such as uncancelled rapidity divergences.
Recently Collins~\cite{Collins:2011zzd} and Echevarria-Idilbi-Scimemi~\cite{GarciaEchevarria:2011rb,Echevarria:2012js} have revisited and updated the definition of quark TMDs, making it consistent with a generic factorization theorem and free from the bad features.
Having at our disposal the proper definition for such quantities allows us to better deal with physical processes where they appear and from which we want to extract sensible information on the hadron structure.
Thus, it is the goal of the present work to extend those efforts to the gluon TMDs, relevant for instance in processes such as Higgs boson and quarkonium production in hadron-hadron collisions.

In order to properly define all the leading-twist (un)polarized gluon TMDPDFs we consider the Higgs boson transverse momentum distribution, generated mainly through the gluon-gluon fusion process. Thus, gluon TMDPDFs will be the relevant hadronic quantities necessary to build our observable. Inclusive Higgs boson production in unpolarized hadron-hadron collisions, has received much attention, both in the context of standard perturbative QCD (see, e.g., \cite{Bozzi:2003jy,Catani:2003zt,Bozzi:2005wk,Bozzi:2007pn,Sun:2011iw,Catani:1988vd,Catani:2000vq,Catani:2010pd,Catani:2011kr}) and soft-collinear effective theory (SCET)~\cite{Bauer:2000yr,Bauer:2001yt,Bauer:2001ct,Beneke:2002ph} (see, e.g., \cite{Chiu:2012ir,Mantry:2009qz,Becher:2012yn,Neill:2015roa}).
TMD gluon correlators were also considered in~\cite{Ji:2005nu,Zhu:2013yxa}.
However none of the previous works paid attention to the cancellation of rapidity divergences in a proper definition of gluon TMDPDFs.
In this paper we reconsider the Higgs $q_T$-distribution in hadron-hadron collisions, but with general polarizations, in order to obtain not only the properly defined unpolarized gluon TMDPDF, but also the polarized ones, i.e., all the leading-twist (un)polarized gluon TMDPDFs.
Their proper definition is crucial in order to be able to address different processes where they are relevant, such as quarkonium (see, e.g.,~\cite{Ma:2012hh,Zhang:2014vmh,Qiu:2013qka}) or heavy-quark pair production (see, e.g.,~\cite{Li:2013mia,Zhu:2013yxa}), and perform consistent phenomenological analyses.

In this work we pay special attention to three of the eight leading-twist gluon TMDPDFs, calculate them explicitly at next-to-leading order (NLO) and demonstrate that the rapidity divergences cancel in their proper definitions.
On one hand, the distributions of unpolarized ($f_{1}^g$) and linearly polarized ($h_{1}^{\perp g}$) gluons inside an unpolarized hadron, and, on the other hand, the gluon helicity TMDPDF ($g_{1L}^g$), which represents the distribution of longitudinally polarized gluons inside a longitudinally polarized hadron. 
The calculation not only supports the definitions introduced in this work, but also allows us to extract valuable perturbative ingredients to resum large logarithms and better control their non-perturbative parts, eventually improving our description of experimental data.
We emphasize that the calculation of $g_{1L}^g$ is done for the first time, while for $f_{1}^g$ and $h_{1}^{\perp g}$ one could combine previous results and then carefully recast them into the new TMD formalism.

The evolution of the gluon TMDPDFs, as in the case of quark TMDs~\cite{Echevarria:2014rua}, turns out to be universal, i.e., the same evolution kernel describes the evolution of any of the leading-twist (un)polarized gluon TMDPDFs. 
It is interesting to contrast this finding with the evolution of the parton distribution functions (PDFs) and double parton distributions (DPDs) which have vastly different evolution depending on the polarization (see, e.g., \cite{Diehl:2014vaa} for a direct comparison for DPDs).
The currently known perturbative ingredients allow us to use the evolution equations to resum the large logarithms up to next-to-next-to-leading-logarithmic (NNLL) accuracy.
Moreover, if we consider the perturbative coefficients of the operator product expansion (OPE) of those TMDs at large transverse momentum, also some parts of them turn out to be universal.
Exploiting this feature, we introduce a further step to resum the large logarithms that appear in the OPE coefficients, exponentiating the double logarithms and improving the convergence of the resummation.
Thus we provide a general framework to deal with (un)polarized gluon TMDPDFs in different processes and account for their perturbative and non-perturbative contributions.

Drawing attention to the distribution of linearly polarized gluons inside an unpolarized hadron, several works have addressed their role at the LHC (see, e.g.,~\cite{Boer:2011kf,Boer:2012bt,Boer:2013fca,Pisano:2013cya,Dunnen:2014eta,Boer:2014tka}).
In particular, in \cite{Boer:2014tka} the authors quantified their contribution in the context of the TMD formalism, both for Higgs boson and $C$-even scalar quarkonium ($\chi_{c0}$ and $\chi_{b0}$) production.
In the present work we extend their efforts by implementing the currently known perturbative ingredients to the full extent to perform the resummation at NNLL accuracy, providing more accurate predictions and discussing their uncertainty.

The paper is organized as follows.
In Section~\ref{sec:factorization} we apply the SCET machinery to derive the factorization theorem for the Higgs $q_T$-distribution in polarized hadron-hadron collisions in terms of well-defined gluon TMDPDFs.
In Section~\ref{sec:evolution} we discuss the QCD evolution of all the leading-twist gluon TMDPDFs, which turns out to be driven by a universal evolution kernel.
Next, in Section~\ref{sec:refactorization} we address the refactorization of TMDPDFs in terms of collinear functions, which applies when the transverse momentum is in the perturbative domain.
In Section~\ref{sec:tmdhelicity} we consider the gluon helicity TMDPDF ($g_{1L}^g$), that accounts for longitudinally polarized gluons inside a longitudinally polarized hadron, and perform a numerical study of the function itself and the impact of evolution.
In Section~\ref{sec:tmdsunpolhadron} we analyze the TMDPDFs that contribute in unpolarized hadron-hadron collisions, i.e., unpolarized and linearly polarized gluons ($f_{1}^g$ and $h_{1}^{\perp g}$ respectively), and give some estimates of their relative contributions at different scales.
Then, in Section~\ref{sec:pheno} we study the Higgs boson transverse momentum distribution, paying special attention to the role played by linearly polarized gluons and the non-perturbative effects. Finally, conclusions are drawn in Section~\ref{sec:conclusions}.

\section{Factorization theorem in terms of well-defined TMDPDFs}
\label{sec:factorization}

Below we derive the factorization theorem for the Higgs $q_T$-distribution in polarized hadron-hadron collisions, $A(P,S_A)+B(\bP,S_B)\to H(m_H,q_T)+X$, by performing a set of consecutive matchings between different effective field theories, relevant at each scale:
$${\rm QCD}(n_f=6)\to{\rm QCD}(n_f=5)\to{\rm SCET}_{q_T}\to{\rm SCET}_{\lqcd}\,.$$
In the first step we integrate out the top quark mass, $m_t$, to build an effective $ggH$ coupling.
In the second matching we integrate out the mass of the Higgs boson, $m_H$, and obtain a factorized cross-section in terms of well-defined gluon TMDPDFs, which holds for $q_T\ll m_H$.
Those gluon TMDPDFs will be expressed in terms of fundamental hadronic matrix elements.
Finally, in the region $\lqcd\ll q_T\ll m_H$, we can further refactorize the gluon TMDPDFs in terms of the collinear gluon/quark PDFs, integrating out the large scale $q_T$.

Before discussing the steps in the derivation of the factorization theorem, we introduce the notation used through the paper.
A generic vector $v^\m$ is decomposed as 
$v^\m=\bn\cd v\frac{n^\m}{2}+n\cd v\frac{\bn^\m}{2}+v_\perp^\m=
(\bn\cd v, n\cd v, \vecb v_\perp)=
(v^+,v^-,\vecb v_\perp)$, with $n=(1,0,0,1)$, $\bn=(1,0,0,-1)$, $n^2=\bn^2=0$ and $n\cd\bn=2$. We also use $v_T=|\vecb v_\perp|$, so that $v_\perp^2=-v_T^2<0$.

The production of the Higgs boson through gluon-gluon fusion is well approximated by the effective local interaction~\cite{Ellis:1975ap,Shifman:1979eb,Vainshtein:1980ea,Inami:1982xt, Voloshin:1985tc}
\begin{align}
{\cal L}_{\rm eff} &= 
C_t(m_t^2,\m)\,\frac{H}{v}\, \frac{\as(\m)}{12\pi} \,
F^{\m\n,a}\,F^a_{\m\n} 
\,,
\end{align} 
where $\alpha_s(\mu)$ is the QCD coupling at factorization scale $\mu$, $F^{\m\n,a}$ the gluon field strength tensor, $H$ is the Higgs field and $v\approx 246$\,GeV is the Higgs vacuum expectation value.
The explicit expressions for the Wilson coefficient $C_t$ and its evolution can be found in Appendix~\ref{app:ads}.
Using the effective lagrangian just introduced, the differential cross section for Higgs production is factorized as
\begin{align}
d\s &=
\frac{1}{2s} \le(\frac{\as(\m)}{12\pi v}\ri)^2 C_t^2(m_t^2,\m)
\frac{d^3q}{(2\pi)^3 2E_q} \int d^4y\, e^{-iq\cd y}
\nn\\
&
\times
\sum_{X} 
\sandwich{PS_A,\bP S_B}{F_{\m\n}^{a} F^{\m\n,a}(y)}{X}
\sandwich{X}{F_{\a\b}^{b} F^{\a\b,b}(0)}{PS_A,\bP S_B}\,,
\end{align}
where $s=(P+\bP)^2$.
This expression manifests the first step in the matching procedure, where we integrate out the large top quark mass through the perturbative coefficient $C_t$.

The $n_f=5$ effective QCD operator is next matched onto the SCET-$q_T$ one:
\begin{align}
F^{\m\n,a}\, F^a_{\m\n} &=
-2q^2 C_H(-q^2,\m^2)\, g^\perp_{\m\n} 
\mB_{n\perp}^{\m,a}
\le( {\cal S}_n^\dagger {\cal S}_{\bn} \ri)^{ab} 
\mB_{\bn\perp}^{\n,b}
\,,
\end{align}
where $q^2=m_H^2$ and the $\mB_{n(\bn)}^{\perp\m}$ operators, which stand for gauge invariant gluon fields, are given by
\begin{align}
\mB_{n\perp}^{\m} &=
\frac{1}{g} [W_n^\dagger iD_n^{\perp\m} W_n] 
=
\frac{1}{\bn\cd{\cal P}}
i\bn_\a g^{\m}_{\perp\b} W_n^\dagger 
F_n^{\a\b} W_n 
= 
\frac{1}{\bn\cd{\cal P}}
i\bn_\a g^\m_{\perp\b} t^a
({\cal W}_n^\dagger)^{ab} F_n^{\a\b,b}
\,.
\end{align}
The collinear and soft Wilson lines are path ordered exponentials
\begin{align}
W_n(x) &=
P\exp\le[ ig\int_{-\infty}^{0} ds\, \bn\cd A_n^a(x+\bn s) t^a\ri]
\,,
\nn\\
S_n(x) &=
P\exp\le[ ig\int_{-\infty}^{0} ds\, n\cd A_s^a(x+n s) t^a\ri]
\,.
\end{align}
Wilson lines with calligraphic typography are in the adjoint representation, i.e., the color generators are given by $(t^a)^{bc}=-if^{abc}$.
In order to guarantee gauge invariance among regular and singular gauges, transverse gauge links need to be added (as described in \cite{Idilbi:2010im,GarciaEchevarria:2011md}).
In this work we stick to Feynman gauge for perturbative calculations, and thus transverse gauge links do not play any role.
The Wilson matching coefficient $C_H(-q^2,\m)$, which corresponds to the infrared finite part of the gluon form factor calculated in pure dimensional regularization, is given at one-loop by
\begin{align}\label{eq:hardcoeff}
C_H(-q^2,\m) &=
1 + \frac{\as C_A}{4\pi}\le[
-\ln^2\frac{-q^2+i0}{\m^2} + \frac{\pi^2}{6}
\ri]
\,.
\end{align}
In Appendix~\ref{app:hard} we report our explicit NLO calculation, and Appendix~\ref{app:ads} gives the evolution and higher order contributions.
We anticipate here that for the phenomenological study discussed in Section~\ref{sec:pheno} we perform the so-called ``$\pi$-resummation'', which consists on choosing the scale in the above coefficient as $\m^2=-q^2$.
In this way the convergence of the hard part is improved.
See~\cite{Magnea:1990zb,Ahrens:2008qu} for more details.

After some standard algebraic manipulations and a Taylor expansion in order to retain the leading order contribution in $q_T/m_H$, the cross-section can be written as
\begin{align}\label{eq:facttheoremjjs}
\frac{d\sigma}{dy\,d^2q_\perp} &= 
2\s_0(\mu)\,C_t^2(m_t^2,\mu) H(m_H,\m)\,
(2\pi)^2 \int d^2\vecb k_{n\perp} d^2\vecb k_{\bn\perp} d^2\vecb k_{s\perp}\,
\d^{(2)}\le( \vecb q_{\perp}-\vecb k_{n\perp}-\vecb k_{\bn\perp}-\vecb k_{s\perp}\ri)
\nn\\
&
\times
J_n^{(0)\m\n}(x_A,\vecb k_{n\perp},S_A;\m)\,
J_{\bn\,\m\n}^{(0)}(x_B,\vecb k_{\bn\perp},S_B;\m)\,
S(\vecb k_{s\perp};\m)
+{\cal O}(q_T/m_H)
 \,,
\end{align}
where $H(m_H^2,\m)=|C_H(-q^2,\m)|^2$, $x_{A,B}=\sqrt{\tau}\,e^{\pm y}$, $\tau=(m_H^2+q_T^2)/s$ and $y$ is the rapidity of the produced Higgs boson. 
The Born-level cross section is
\begin{equation}
\s_0(\mu) = \frac{m_H^2\,\as^2(\mu)}{72\pi (N_c^2-1) s v^2} 
\,.
\end{equation}
The \emph{pure} collinear matrix elements and the soft function are defined as
\begin{align}
J_n^{(0)\m\n}(x_A,\vecb k_{n\perp},S_A;\m) &= 
\frac{x_A P^+}{2} \int\frac{dy^-d^2\vecb y_\perp}{(2\pi)^3}\,
e^{-i\le(\frac{1}{2}x_A y^-P^+-\vecbe y_\perp\cd\vecbe k_{n\perp}\ri)}
\nn\\ 
&\times
\sum_{X_n}\,
\sandwich{PS_A}{B_{n\perp}^{\m,a}(y^-,\vecb y_\perp)}{X_n}\,
\sandwich{X_n}{B_{n\perp}^{\n,a}(0)}{PS_A}
\,,\nn\\
J_\bn^{(0)\m\n}(x_B,\vecb k_{\bn\perp},S_B;\m) &= 
\frac{x_B \bP^-}{2} \int\frac{dy^+d^2\vecb y_\perp}{(2\pi)^3}\,
e^{-i\le(\frac{1}{2}x_B y^+\bP^--\vecbe y_\perp\cd\vecbe k_{\bn\perp}\ri)}
\nn\\ 
&\times
\sum_{X_\bn}\,
\sandwich{\bP S_B}{B_{\bn\perp}^{\m,a}(y^+,\vecb y_\perp)}{X_\bn}\,
\sandwich{X_\bn}{B_{\bn\perp}^{\n,a}(0)}{\bP S_B}
\,,\nn\\
S(\vecb k_{s\perp};\m) &= 
\frac{1}{N_c^2-1}\,\sum_{X_s}\,
\int\frac{d^2\vecb y_\perp}{(2\pi)^2}\,
e^{i\vecbe y_\perp\cd\vecbe k_{s\perp}}
\sandwich{0}{\big({\cal S}_n^\dagger{\cal S}_{\bar n}\big)^{ab}(\vecb y_\perp)}{X_s}\,
\sandwich{X_s}{\big({\cal S}_{\bar n}^\dagger{\cal S}_n\big)^{ba}(0)}{0}
\,.
\end{align}
Notice that, in order to avoid double counting, one needs to subtract the contribution of soft momentum modes (the so-called ``zero-bin''  in the SCET nomenclature) from the naively calculated collinear matrix elements, thus obtaining the ``pure collinear'' matrix elements, denoted by the superscript $(0)$ (see, e.g., \cite{Manohar:2006nz} for more details).
Note that in \eq{eq:facttheoremjjs} we have applied the SCET machinery to decouple the collinear, anticollinear and soft modes, removing the interactions between them from the Lagrangian~\cite{Bauer:2001yt}. 
Thus the factorization of any operator in SCET is straighforward, as well as the final states $\ket{X}$, which can be written in the factorized form $\ket{X}=\ket{X_n}\otimes\ket{X_\bn}\otimes\ket{X_s}$, describing the collinear, anticollinear and soft states.

As shown in Appendices \ref{app:tmdpdfnlo}, \ref{app:linearnlo} and \ref{app:helicitynlo} by performing an explicit NLO perturbative calculation of unpolarized, linearly polarized and helicity gluon TMDPDFs, the collinear and soft matrix elements defined above are individually ill-defined, since they contain uncancelled rapidity divergences.
We stress the fact that this issue is independent of the particular regulator used.
Thus, we need to combine them in a certain way to cancel these divergences and obtain well-defined hadronic quantities.
Based on the work done in \cite{GarciaEchevarria:2011rb,Echevarria:2012js,Echevarria:2014rua} for the quark case and using $\eta_{n(\bn)}$ to label generic parameters that regulate the rapidity divergences present in the (anti-)collinear and soft matrix elements, then the TMDPDFs are defined as
\begin{align}\label{eq:tmdsdefinition0}
{\tilde G}_{g/A}^{\m\n}(x_A,\vecb b_{\perp},S_A;\z_A,\m) &=
\tilde{J}_{n}^{(0)\m\n}(x_A,\vecbp b,S_A;\m;\eta_n)\,
\tilde{S}_-(b_T;\m;\eta_n)
\,,
\nn\\
{\tilde G}_{g/B}^{\m\n}(x_B,\vecb b_{\perp},S_B;\z_B,\m) &=
\tilde{J}_{\bn}^{(0)\m\n}(x_B,\vecbp b,S_B;\m;\eta_\bn)\,
\tilde{S}_+(b_T;\m;\eta_\bn)
\,,
\end{align}
where $\z_{A,B}$ are auxiliary energy scales, the twiddle labels the functions in impact parameter space (IPS) and we have split the soft function in rapidity space as
\begin{align}\label{eq:softsplitting}
\tilde{S}(b_T;\m;\eta_n,\eta_\bn) &=
\tilde{S}_{-}\le(b_T;\m;\eta_n\ri)
\tilde{S}_{+}\le(b_T;\m;\eta_\bn\ri)
\,.
\end{align}
The soft function can be split to all orders in perturbation theory, regardless which particular regulator is used, following the same logic as in \cite{Echevarria:2012js}.
In that work this fundamental property was proven for the soft function relevant for quark TMDs.
The proof for the soft function that appears in the gluon TMDs follows analogously, simply changing the color representation from the fundamental to the adjoint.
The arbitrariness in the choice of the rapidity cutoff to split the soft function, which is not explicitly shown in \eq{eq:softsplitting}, manifests itself as the appearance of the auxiliary energy scales $\z_A$ and $\z_B$, which are bound together by $\z_A\z_B=q^4=m_H^4$.

Resorting to the $\D$-regulator for definiteness, the soft function is split as
\begin{align}\label{eq:splittingdelta}
\tilde{S}(b_T;m_H^2,\m; \D^+,\D^-) &=
\tilde{S}_{-}\le(b_T;\z_A,\m;\D^-\ri)
\tilde{S}_{+}\le(b_T;\z_B,\m;\D^+\ri)
\,,
\nn\\
\tilde{S}_{-}\le(b_T;\z_A,\m;\D^-\ri) &=
\sqrt{\tilde S\left(\frac{\D^-}{p^+},\a\frac{\D^-}{\bar {p}^-}\right)}
\,,
\nn\\
\tilde{S}_{+}\le(b_T;\z_B,\m;\D^+\ri) &=
\sqrt{\tilde S\left(\frac{1}{\a}\frac{\D^+}{p^+},\frac{\D^+}{\bar {p}^-}\right)}
\,,
\end{align}
where we have explicitly shown the dependence on the regulator parameters and $\z_A=m_H^2/\a$ and $\z_B=\a m_H^2$, with $\a$ an arbitrary boost invariant real parameter.

We point out that an explicit dependence on $q^2=m_H^2$ has been added as well in the soft function.
This is due to the use of the $\D$-regulator, which induces such dependence, in consistency with the fact that the soft function represents the cross-talking between the two collinear sectors.
On the contrary, pure collinear matrix elements do not have any remnant information about the opposite collinear sector, and thus cannot depend on $q^2=m_H^2$. 

Now, using \eq{eq:splittingdelta} the gluon TMDPDFs are defined as
\begin{align}\label{eq:tmdsdefinition}
{\tilde G}_{g/A}^{\m\n}(x_A,\vecb b_{\perp},S_A;\z_A,\m;\D^-) &=
\tilde{J}_{n}^{(0)\m\n}(x_A,\vecbp b,S_A;\m;\D^-)\,
\tilde{S}_-(b_T;\z_A,\m;\D^-)
\,,
\nn\\
{\tilde G}_{g/B}^{\m\n}(x_B,\vecb b_{\perp},S_B;\z_B,\m;\D^+) &=
\tilde{J}_{\bn}^{(0)\m\n}(x_B,\vecbp b,S_B;\m;\D^+)\,
\tilde{S}_+(b_T;\z_B,\m;\D^+)
\,.
\end{align}
These hadronic quantities are free from rapidity divergences, i.e., they have well-behaved evolution properties and can be extracted from experimental data.

As already mentioned, when one performs the perturbative calculations of the collinear matrix elements, depending on the particular regulator that is used, the issue of double counting the soft modes arises.
In this case, one needs to subtract, on a diagram-by-diagram basis, the soft limit of each collinear contribution (the ``zero-bin'').
With the $\D$-regulator one can show order by order in perturbation theory that the subtraction of the zero-bin for each collinear matrix element is equivalent to divide it by the soft function:
\begin{align}\label{eq:tmdsdefinition2}
{\tilde J}_{n}^{(0)\m\n}(x_A,\vecbp b,S_A;\m;\D^-) &=
\frac{{\tilde J}_{n}^{\m\n}(x_A,\vecbp b,S_A;m_ H^2,\m^2;\D^-,\D^+)}
{\tilde{S}(b_T;m_H^2,\m;\D^+,\D^-)}
\,.
\end{align}
The naively calculated collinear matrix elements, which do not have the label $(0)$ anymore, depend on the hard scale $q^2=m_H^2$ and the two regulator parameters $\D^\pm$, in contrast with the \emph{pure} collinear matrix elements.
The latter should depend only on the regulator that belongs to each collinear sector.
The naive collinear matrix elements depend as well on the spurious regulator for the other sector.
Thus, for the particular case of this regulator we can define the TMDPDFs as
\begin{align}\label{eq:tmdsdefinition2}
{\tilde G}_{g/A}^{\m\n}(x_A,\vecb b_{\perp},S_A;\z_A,\m;\D^-) &=
{\tilde J}_{n}^{\m\n}(x_A,\vecbp b,S_A;m_ H^2,\m;\D^-,\D^+)\,
\tilde{S}_+^{-1}(b_T;\z_B,\m;\D^+)
\,,
\nn\\
{\tilde G}_{g/B}^{\m\n}(x_B,\vecb b_{\perp},S_B;\z_B,\m;\D^+) &=
{\tilde J}_{\bn}^{\m\n}(x_B,\vecbp b,S_B;m_H^2,\m;\D^-,\D^+)\,
\tilde{S}_-^{-1}(b_T;\z_A,\m;\D^-)
\,.
\end{align}
Notice that the spurious regulator in the naive collinear matrix elements is now cancelled by dividing them by the proper piece of the soft function, thus recovering the correct regulator dependence in the TMDPDFs as in Eq.~\eqref{eq:tmdsdefinition}.

It is worth emphasizing that the proper definition of TMDPDFs in Eq.~\eqref{eq:tmdsdefinition0} is independent of the particular regulator used.
We have given their definition using the $\D$-regulator, although one could conveniently modify Eq.~\eqref{eq:tmdsdefinition} in order to use, for instance, the rapidity regulator introduced in \cite{Chiu:2011qc}.
One could also follow the lines of \cite{Collins:2011zzd}, where no regulator is used, i.e., the rapidity divergences among the collinear and soft matrix elements are cancelled by combining (before integration) the integrands of the relevant Feynman diagrams order by order in perturbation theory.

Now, we can write the cross-section for the Higgs $q_T$-distribution in terms of well-defined gluon TMDPDFs:
\begin{align}\label{eq:factth}
\frac{d\sigma}{dy\,d^2q_\perp} &= 
2\s_0(\mu)\,C_t^2(m_t^2,\mu) H(m_H^2,\m)\,
\frac{1}{(2\pi)^2} \int d^2y_\perp\,e^{i\vecbe q_\perp\cd\vecbe y_\perp} 
\nn\\
&
\times
{\tilde G}_{g/A}^{\m\n}(x_A,\vecb y_\perp,S_A;\z_A,\m)\,
{\tilde G}_{g/B\,\m\n}(x_B,\vecb y_\perp,S_B;\z_B,\m)
+{\cal O}(q_T/m_H)
\,.
\end{align}
This factorized cross-section is valid for $q_T\ll m_H$.
In the next section we discuss the refactorization of the TMDPDFs in terms of collinear functions when the transverse momentum is a perturbative scale, i.e., when $\lqcd\ll q_T\ll m_H$.

Before moving to the next factorization step, we first need to consider the dependence on the hadron spin and separate the unpolarized (U),
longitudinally polarized (L) and transversely polarized (T) situations.
In \cite{Mulders:2000sh} the authors obtained the decomposition of collinear correlators at leading-twist.
We emphasize the fact that these correlators suffer from rapidity divergences and thus cannot be considered well-defined hadronic quantities.
The decomposition, however, is not directly affected by this issue and follows equivalently.
Below, given the proper definition of gluon TMDPDFs in Eq.~\eqref{eq:tmdsdefinition0}, we extend the decomposition of \cite{Mulders:2000sh} and write
\begin{align}\label{eq:decomposition}
G_{g/A}^{\m\n[U]} (x_A,\vecb k_{n\perp}) &= 
-\frac{g_\perp^{\m\n}}{2}f_1^g(x_A,k_{nT})
+ \frac{1}{2}\le(g_\perp^{\m\n} - \frac{2k_{n\perp}^\m k_{n\perp}^\n}{k_{n\perp}^2}\ri) h_1^{\perp g}(x_A,k_{nT})
\,,\nn\\
G_{g/A}^{\m\n[L]} (x_A,\vecb k_{n\perp}) &=
-i\frac{\e^{\m\n}_\perp}{2} \l\, g_{1L}^g(x_A,k_{nT})
+ \frac{\e_\perp^{k_{n\perp}\{\m}k_{n\perp}^{\n\}}}{2k_{n\perp}^2}
\l\, h_{1L}^{\perp g}(x_A,k_{nT})
\,,\nn\\
G_{g/A}^{\m\n[T]} (x_A,\vecb k_{n\perp}) &=
-g_\perp^{\m\n} \frac{\e_\perp^{k_{n\perp} S_\perp}}{k_{nT}} 
f_{1T}^{\perp g}(x_A,k_{nT})
-i\e_\perp^{\m\n} \frac{\vecb k_{n\perp}\cd\vecb S_\perp}{k_{nT}} 
g_{1T}^g(x_A,k_{nT}) 
\nn\\
&
+ \frac{\e_\perp^{k_{n\perp}\{\m}k_{n\perp}^{\n\}}}{2k_{n\perp}^2}
\frac{\vecb k_{n\perp}\cd\vecb S_\perp}{k_{nT}} h_{1T}^{\perp g}(x_A,k_{nT}) 
+ \frac{\e_\perp^{k_{n\perp}\{\m}S_\perp^{\n\}}+\e_\perp^{S_\perp\{\m}k_{n\perp}^{\n\}}}{4k_{nT}}
h_{1T}^{g}(x_A,k_{nT}) 
\,.
\end{align}
The functions $f_1^g$, $h_1^{\perp g}$, $g_{1L}^{g}$ and $g_{1T}^{g}$ are $T$-even, while the rest are $T$-odd.
In Sections~\ref{sec:tmdsunpolhadron} and~\ref{sec:tmdhelicity} we will pay special attention to the functions $f_1^g$, $h_1^{\perp g}$ and $g_{1L}^{g}$, calculating them explicitly at NLO to show that they are free from rapidity divergences and to obtain the necessary perturbative ingredients to perform the resummation of large logarithms. 
These three functions are the only TMDPDFs which are matched onto leading twist collinear matrix elements, i.e., the canonical PDFs. 

The Wilson line structure in the operator definition of the TMDPDFs gives rise to calculable process dependence. In the types of processes considered here, where two gluons fuse into a color singlet, the Wilson lines are past pointing. In a process where color would flow into the final state, also future pointing Wilson lines play a role. Functions with different Wilson line structure differ by matrix elements containing intrinsically nonlocal gluonic pole contributions~\cite{Efremov:1981sh,Efremov:1984ip,Qiu:1991pp,Qiu:1991wg}. Depending on the number of such gluonic poles being odd or even, the functions are T-odd or T-even. The functions come with specific process-dependent gluonic pole factors that can lead to a breaking of universality, in the simplest cases giving rise to a sign change, such as the Sivers function having a different sign in Drell-Yan and in deep-inelastic scattering (DIS) processes.  
Other functions, such as $h_1^{\perp g}$ need to be written as a linear combination of two or even more functions, with the coefficient in the linear combination depending on the Wilson lines and in turn on the color flow in the process~\cite{Bomhof:2007xt,Buffing:2013dxa}. 
However, for both the gluon-gluon fusion into a color singlet considered here, as well as in the ``gluon initiated DIS'', exactly the same linear combination contributes to the cross section. 

Finally, we provide the equivalent of Eq.~\eqref{eq:decomposition} in IPS, since as we show next, the evolution of TMDPDFs is done in that space.
With the Fourier transform given by
\begin{align}
{\tilde G}_{g/A}^{\m\n[pol]} (x_A,\vecb b_{\perp}) &=
\int d^2\vecb k_{n\perp}\,
e^{i\vecbe k_{n\perp}\cd \vecbe b_{\perp}}\,
G_{g/A}^{\m\n[pol]} (x_A,\vecb k_{n\perp})
\,,
\end{align}
we have
\begin{align}\label{eq:decompositionips}
{\tilde G}_{g/A}^{\m\n[U]} (x_A,\vecb b_{\perp}) &= 
-\frac{g_\perp^{\m\n}}{2}{\tilde f}_1^g(x_A,b_{T})
+ \frac{1}{2}\le(g_\perp^{\m\n} - \frac{2b_{\perp}^\m b_{\perp}^\n}{b_{\perp}^2}\ri) 
{\tilde h}_1^{\perp g\,(2)}(x_A,b_{T})
\,,\nn\\
{\tilde G}_{g/A}^{\m\n[L]} (x_A,\vecb b_{\perp}) &=
-i\frac{\e^{\m\n}_\perp}{2} \l\, {\tilde g}_{1L}^g(x_A,b_{T})
+ \frac{\e_\perp^{b_\perp\{\m}b_{\perp}^{\n\}}}{2b_\perp^2}
\l\, h_{1L}^{\perp g\,(2)}(x_A,b_{T})
\,,\nn\\
{\tilde G}_{g/A}^{\m\n[T]} (x_A,\vecb b_{\perp}) &=
-g_\perp^{\m\n} \frac{\e_\perp^{b_{\perp} S_\perp}}{b_T} 
{\tilde f}_{1T}^{\perp g\,(1)}(x_A,b_{T})
-i\e_\perp^{\m\n} \frac{\vecb b_{\perp}\cd\vecb S_\perp}{b_T} 
{\tilde g}_{1T}^{g\,(1)}(x_A,b_{T}) 
\nn\\
&
+ \frac{\e_\perp^{b_\perp\{\m}b_{\perp}^{\n\}}}{2b_\perp^2}
\frac{\vecb b_{\perp}\cd\vecb S_\perp}{b_T} 
{\tilde h}_{1T}^{\perp g\,(2)}(x_A,b_{T}) 
+ \frac{\e_\perp^{b_\perp\{\m}S_\perp^{\n\}}
+\e_\perp^{S_\perp\{\m}b_{\perp}^{\n\}}}{4b_T}
{\tilde h}_{1T}^{g\,(1)}(x_A,b_{T}) 
\,,
\end{align}
where for a generic function $f(k_T)$ we represent
\begin{align}\label{eq:moments}
{\tilde f}^{(n)}(b_T) &=
2\pi (i)^n \int dk_T k_T\, J_n(k_T b_T)\, f(k_T)
\,.
\end{align}
Worth noting is that while $G_{g/A}^{\m\n}$ and ${\tilde G}_{g/A}^{\m\n}  $ are each others Fourier transforms, this does not hold true for the individual gluon TMDs which have factors of $k_{n\perp}$ ($b_\perp$) in the decomposition (e.g., $h_1^{\perp g}$ and ${\tilde h}_{1T}^{g\,(1)}$).
\section{Evolution of Gluon TMDPDFs}
\label{sec:evolution}

The TMDPDFs defined in Eq.~\eqref{eq:tmdsdefinition0} depend on two scales: the factorization scale $\mu$ and the energy scale $\z$ (related to the rapidity cutoff used to separate the two TMDPDFs).
Thus, their evolution kernel is such that it connects these two scales between their initial and final values.
Below we derive first the part of the kernel that allows us to evolve the TMDPDFs with respect to $\m$, and then the one that corresponds to $\z$.

The evolution of (un)polarized gluon TMDPDFs in terms of the renormalization scale $\m$ is governed by the anomalous dimensions:
\begin{align}
\frac{d}{d\ln\m}\ln\tilde{G}^{[pol]}_{g/A}(x_A,\vecb b_\perp,S_A;\z_A,\m) &\equiv
\g_G\le(\as(\m),\ln\frac{\z_A}{\m^2}\ri)
\,,
\nn\\
\frac{d}{d\ln\m}\ln\tilde{G}^{[pol]}_{g/B}(x_B,\vecb b_\perp,S_B;\z_B,\m) &\equiv
\g_G\le(\as(\m),\ln\frac{\z_B}{\m^2}\ri)
\,.
\end{align}
The renormalization group (RG) equation applied to the factorized cross-section in Eq.~\eqref{eq:factth} implies the following relation among the different anomalous dimensions:
\begin{align}
2\frac{\b\le(\as(\m)\ri)}{\as(\m)} 
+ 2\g^t\le(\as(\m)\ri) 
+ \g_H\le(\as(\m),\ln\frac{m_H^2}{\m^2}\ri)
+ \g_G\le(\as(\m),\ln\frac{\z_A}{\m^2}\ri)
+ \g_G\le(\as(\m),\ln\frac{\z_B}{\m^2}\ri)
&= 0
\,,
\end{align}
where the anomalous dimension of the coefficients $H$ and $C_t$, $\g_H$ and $\g^t$ respectively, are given in Appendix~\ref{app:ads}.
Thus
\begin{align}\label{eq:gammag}
\g_G\le(\as(\m),\ln\frac{\z_A}{\m^2}\ri) &=
-\G_{\rm cusp}^A(\as(\m))\ln\frac{\z_A}{\m^2} - \g^{nc}(\as(\m))
\,,
\nn\\
\g_G\le(\as(\m),\ln\frac{\z_B}{\m^2}\ri) &=
-\G_{\rm cusp}^A(\as(\m))\ln\frac{\z_B}{\m^2} - \g^{nc}(\as(\m))
\,,
\end{align}
where the non-cusp piece is
\begin{align}
\g^{nc}(\as(\m)) &=
\g^g(\as(\m)) + \g^t(\as(\m)) + \frac{\b(\as(\m))}{\as(\m)}
\,.
\end{align}
In the equation above $\g^{g}$ is the non-cusp piece of the anomalous dimension of the hard coefficient $C_H$ (see Appendix~\ref{app:ads}). 
It should be mentioned that the splitting of $\g_H$ into the two anomalous dimensions $\g_G$ given in \eq{eq:gammag} is unique following the restriction $\z_A\z_B=m_H^4$. 
The coefficients of the perturbative expansions of $\G_{\rm cusp}$ and $\g^V$ are known up to three loops and they are collected in Appendix~\ref{app:ads}.

Now we focus our attention on the evolution in terms of the scale $\z$.
Following the arguments in \cite{Echevarria:2012js}, one can show that the soft function relevant for gluon TMDs can to all orders be written as
\begin{align}\label{eq:splitting}
\ln{\tilde S} &=
{\cal R}_{s}(b_T;\m) + D_g(b_T;\m)\, \ln\frac{\D^+\D^-}{m_H^2\m^2}
\,,
\end{align}
with a function ${\cal R}_s$, depending only on $b_T$ and $\m$, and $D_g$ related to the cusp anomalous dimension in the adjoint representation by
\begin{align}\label{eq:devol}
\frac{dD_g}{d\ln\m} &=
\G_{\rm cusp}^A(\as(\m))
\,.
\end{align}
Given Eqs.~\eqref{eq:tmdsdefinition} and~\eqref{eq:splitting}, one obtains the following evolution equations in $\z$:
\begin{align}\label{eq:evolzeta}
\frac{d}{d\ln\z_A}\ln\tilde{G}^{[pol]}_{g/A}(x_A,\vecb b_\perp,S_A;\z_A,\m) &=
- D_g(b_T;\m)
\,,
\nn\\
\frac{d}{d\ln\z_B}\ln\tilde{G}^{[pol]}_{g/B}(x_B,\vecb b_\perp,S_B;\z_B,\m) &=
- D_g(b_T;\m)
\,.
\end{align}
Notice that the same $D_g$ term drives the $\z$ evolution for all gluon TMDPDFs, since the soft function that enters into their definition and gives the entire $\z$ evolution is spin-independent.
The coefficients of the perturbative expansion of the $D_g$ term can be completely obtained from the calculation of the soft function.
If we write
\begin{align}
D_g(b_T;\m) &=
\sum_{n=1}^{\inf} d_n(L_T) \le(\frac{\as(\m)}{4\pi}\ri)^n
\,,\quad\quad
L_T = \ln\frac{\m^2 b_T^2}{4e^{-2\g_E}}
\,,
\end{align}
then the first two coefficients are:
\begin{align}\label{eq:dgcoeffs}
d_1(L_T) &=
\frac{\G_0^A}{2\b_0} (\b_0 L_T) + d_1(0)
\,,\nn\\
d_2(L_T) &=
\frac{\G_0^A}{4\b_0} (\b_0 L_T)^2 + \le(\frac{\G_1^A}{2\b_0}+d_1(0)\ri) 
+ d_2(0)
\,.
\end{align}
The finite coefficients $d_n(0)$ cannot be determined by Eq.~\eqref{eq:devol}, but, as already mentioned, by a perturbative calculation of the soft function (or the cross-section in full QCD).
The coefficient $d_1(0)$ can be easily extracted from the NLO calculation of the soft function in Appendix~\ref{app:tmdpdfnlo}, and one gets $d_1(0)=0$.
The coefficient $d_2(0)$ can be obtained from the soft function relevant for DY or SIDIS processes~\cite{Echevarria:2012pw} by using the Casimir scaling, i.e., rescaling it by $C_A/C_F$ (see also~\cite{Becher:2012yn}):
\begin{align}
d_2(0) &=
C_A C_A \left(\frac{404}{27}-14\z_3\right)
- \left(\frac{112}{27}\right)C_A T_F n_f
\,.
\end{align}
At small $b_T$ the $D_g$ term can be calculated perturbatively, but at large $b_T$ it has to be modelled and extracted from experimental data.
However, we can extend the fact that the soft function is universal and spin-independent to this non-perturbative piece, and it can therefore be used to parametrize the non-perturbative contribution to the evolution of all (un)polarized TMDPDFs.

Regardless of how the non-perturbative contribution to the $D_g$ term is parametrized, we can perform the evolution of all leading-twist gluon TMDPDFs consistently up to NNLL (given the currently known perturbative ingredients, i.e., $\g_G$ and $D_g$):
\begin{align}
{\tilde G}^{[pol]}_{g/A}(x_A,\vecb b_\perp,S_A;\z_{A,f},\m_f) &=
{\tilde G}^{[pol]}_{g/A}(x_A,\vecb b_\perp,S_A;\z_{A,i},\m_i)\,
{\tilde R}^g\le(b_T;\z_{A,i},\m_i,\z_{A,f},\m_f\ri)
\,,
\end{align}
where the evolution kernel $\tilde{R}^g$ is given by
\begin{align}\label{eq:evolkernel}
{\tilde R}^g\big(b_T;\z_{A,i},\m_i,\z_{A,f},\m_f\big) &=
\exp\le\{
\int_{\m_i}^{\m_f} \frac{d\bar\m}{\bar\m}\, 
\g_G\le(\as(\bar\m),\ln\frac{\z_{A,f}}{\bar\m^2} \ri)
\ri\}
\le( \frac{\z_{A,f}}{\z_{A,i}} \ri)^{-D_g\le(b_T;\m_i\ri)}
\,.
\end{align}

Solving analytically the evolution equation of the $D_g$ term in the small $b_T$ region,
\begin{align}
D_g^R(b_T;\m_i) &=
D_g(b_T;\m_b) + \int_{\m_b}^{\m_i}\frac{d\bar\m}{\bar\m} \G^A_{\rm cusp}
\,,
\end{align}
where $\m_b=2e^{-\g_E}/b_T$ is the natural scale of the $D_g$ term, and implementing the running of the strong coupling consistently with the resummation order, one obtains (see~\cite{Echevarria:2012pw} for quark TMDs)
\begin{align}\label{eq:dr}
D_g^R(b_T;\m_i) &=
-\frac{\Ga_0^A}{2\beta_0}\ln(1-X)
+ \frac{1}{2}\le(\frac{a}{1-X}\ri) \le[
- \frac{\beta_1\Ga_0^A}{\beta_0^2} (X+\ln(1-X))
+\frac{\Ga_1^A}{\beta_0} X\ri]
\nn\\
&+ \frac{1}{2}
\le(\frac{a}{1-X}\ri)^2\le[
2d_2(0)
+\frac{\Ga_2^A}{2\beta_0}(X (2-X
))
+\frac{\beta_1\Ga_1^A}{2 \beta_0^2} \le( X (X-2)-2 \ln (1-X)\ri) 
\ri.
\nn\\
&\le.
+\frac{\beta_2\Ga_0^A}{2\beta_0^2} X^2
+\frac{\beta_1^2\Ga_0^A}{2 \beta_0^3} (\ln^2(1-X)-X^2)
\ri]
 +\mathcal{O}\left(\left(\frac{a}{1-X}\right)^3\right)
\,.
\end{align}
In this result we have defined $a=\as(\m_i)/(4\pi)$ and $X=a\beta_0 L_T$.
The $\beta_i$ and $\Gamma_i^A$ coefficients are given in Appendix~\ref{app:ads}.

As a final remark, we emphasize the fact that the evolution kernel in \eq{eq:evolkernel} is valid only in the perturbative region of small $b_T \lesssim \lqcd^{-1}$, since the perturbative expression of $D_g$ (even its resummed version) breaks down at large $b_T$~\cite{Echevarria:2012pw,Collins:2014jpa}.

\section{Refactorization of TMDPDFs and Resummation of Large Logarithms}
\label{sec:refactorization}

As already anticipated, when the transverse momentum is perturbative, we can perform an operator product expansion (OPE) of the TMDPDFs in terms of collinear functions, integrating out the transverse momentum by means of Wilson coefficients.
Depending on the particular TMDPDF considered, the collinear functions that will describe its perturbative small-$b_T$ region will be different, and also the relevant Wilson coefficients.
However, the part of the Wilson coefficients originating from the evolution, which is universal and spin-independent, will be common for all TMDPDFs.
Below we give the general expressions for these OPEs, and in the next sections we explicitly calculate the coefficients for the relevant TMDPDFs in the cases of an unpolarized and longitudinally polarized hadron.

For $b_T\ll \lqcd^{-1}$ we can refactorize the (renormalized) gluon TMDPDFs of a hadron $A$ in terms of (renormalized) collinear quark/gluon distributions:
\begin{align}
{\tilde F}_{g/A}(x_A,b_T;\z_A,\m) &=
\sum_{j=q,\bar q, g}
{\tilde C}_{g/j}(x_A,b_T;\z_A,\m)\otimes f_{j/A}(x_A;\m)
+{\cal O}(b_T\lqcd)
\,.
\end{align}
The convolution refers to momentum fraction $x$ for TMDPDFs that are matched onto twist-2 collinear functions (like $f_1^g$), while in the case of TMDPDFs that are matched onto twist-3 functions (like the gluon Sivers function $f_{1T}^{\perp g}$) it would represent a two-dimensional convolution in the two momentum fractions of the collinear function. In the equation above we have thus represented schematically the OPE of any TMDPDF, where ${\tilde F}_{g/A}$ stands for any of the functions in Eq.~\eqref{eq:decompositionips} and $f_{j/A}$ the adequate collinear functions in each case. 
For example, we could consider the unpolarized gluon TMDPDF ${\tilde f}_1^g$ and match it onto the unpolarized collinear gluon/quark PDFs, as shown in Section~\ref{sec:tmdsunpolhadron};
or we could consider the Sivers function ${\tilde f}_{1T}^{\perp g\,(1)}$ and match in onto gluon/quark twist-3 collinear functions~\cite{Ma:2012xn}.
The coefficients ${\tilde C}_{g/j}$ are different for each TMDPDF.

The natural scale for the coefficients ${\tilde C}_{g/j}$ is $\m\sim 1/b_T\sim q_T$, which is the large scale that we integrate out when we perform the OPE.
Thus, we can choose to set the resummation scale either in impact parameter space or in momentum space.
In the following we discuss these two approaches in more detail.

\subsection{Resummation in Impact Parameter Space}
\label{sec:impact}

If we perform the resummation of large logarithms in impact parameter space then the resummed TMDPDF is written as:
\begin{align}
{\tilde F}_{g/A}^{Pert}(x_A,b_T;\z_A,\m) &=
\exp\le\{
\int_{\m_0}^{\m} \frac{d\bar\m}{\bar\m}
\g_G\le(\as(\bar\m),\ln\frac{\z_A}{\bar\m^2} \ri)\ri\}\,
\le(\frac{\z_A}{\z_0}\ri)^{-D_g(b_T;\m_0)}
\nn\\
&\times
\sum_{j=q,\bar q, g}
{\tilde C}_{g/j}(x_A,b_T;\z_0,\m_0)\otimes
f_{j/A}(x_A;\m_0)\,
\,,
\end{align}
where $\z_0\sim \m_b^2$ and $\m_0\sim \m_b$.
The superscript $Pert$ signifies that it is only the perturbative part of the TMDPDFs~\footnote{We refer to the perturbative or non-perturbative nature of the transverse momentum (or impact parameter) dependence, leaving aside of course the non-perturbative collinear distributions, which are always part of the OPE, both in the small and large $b_T$ regions.}, valid at small $b_T<<1/\Lambda_{QCD}$.
Notice that the functions $D_g$ is universal and spin independent, and thus it is the same for any of the TMDPDFs in Eq.~\eqref{eq:decompositionips}.
On the contrary, as already mentioned, the coefficients ${\tilde C}_{g/j}$ are specific for each TMDPDF, as are the collinear functions $f_{j/A}$ which generate the perturbative tail for each TMDPDF at small $b_T$.

So far we have addressed the TMDPDFs in the perturbative region.
For large $b_T$ we need to model them and extract them from experimental data.
To do so, one could implement a smooth cutoff that freezes the perturbative contribution slowly as $b_T$ gets larger:
\begin{align}\label{eq:nonp-bhat}
{\tilde F}_{g/A}(x_A,b_T;\z_A,\m) &= 
{\tilde F}_{g/A}^{Pert}(x_A,{\hat b}_T;\z_A,\m)\,
{\tilde F}_{}^{NP}(x_A,b_T;\z_A)
\,,
\end{align}
where the cutoff prescription could be, for instance:
\begin{align}
\hat b_T(b_T) = b_c\le( 1 - e^{-(b_T/b_c)^n} \ri)^{1/n}
\,,
\end{align}
with $n$ an integer number and $b_c$ the parameter that determines the separation between the perturbative and non-perturbative regions.
For small $b_T$ the perturbative contribution dominates, and gets frozen as we increase the $b_T$, since ${\hat b}_T\to b_{c}$ for large $b_T$.
The non-perturbative model ${\tilde F}^{NP}$ is constrained to be $1$ for $b_T=0$ and plays an increasingly important role as we increase $b_T$.

\subsection{Resummation in Momentum Space}
\label{sec:momentum}

Instead of setting $\m_0\sim\m_b$, in this case we keep it in momentum space.
In this way we avoid hitting the Landau pole in the strong coupling, and we write the TMDPDF as
\begin{align}
{\tilde F}_{g/A}^{Pert}(x_A,b_T;\z_A,\m) &=
\exp\le\{
\int_{\m_0}^{\m} \frac{d\bar\m}{\bar\m}
\g_G\le(\as(\bar\m),\ln\frac{\z_A}{\bar\m^2} \ri)\ri\}\,
\le(\frac{\z_A}{\z_0}\ri)^{-D_g(b_T;\m_0)}
\nn\\
&\times
\sum_{j=q,\bar q, g}
{\tilde C}_{g/j}(x_A,b_T;\z_0,\m_0)\otimes
f_{j/A}(x_A;\m_0)\,
\,,
\end{align}
where $\z_0=C_\z^2 \m_b^2$ and $\m_0\sim q_T$.
We have kept explicitly the dependence on the real parameter $C_\z$, which will be used later on to test the dependence of the results on the rapidity scale, basically varying it between $1/2$ and $2$.
This is because the way it enters in the final resummed expression for the TMDPDF is subtle, contrary to the scale $\m_0$, which can be easily identified where it appears.

Now the coefficients ${\tilde C}_{g/j}$ contain large logarithms $\ln(\m_0 b_T)$ which are not minimized by the choice $\m_0\sim q_T$ (this was the case in the previous subsection, when choosing $\m_0\sim 1/b_T$).
However we can further split them by using their RG-equation
\begin{align}
\frac{d}{d\ln \mu} \tilde{C}_{g/j}(x,b_T;C_\z^2\m_b^2,\m) &=
(\G_{\rm cusp}^A L_T - \g^{nc} - \G_{\rm cusp}^A\ln C_\z^2)
\tilde{C}_{g/j}(x,b_T;C_\z^2\m_b^2,\m)
\nn\\
&
-\sum_i\int_x^1\frac{dz}{z}
\tilde{C}_{g/i}(z,b_T;C_\z^2\m_b^2,\m)\, {\cal P}_{i/j}(x/z)
\,,
\end{align}
where ${\cal P}_{i/j}(x/z)$ are the usual DGLAP splitting kernels, so that double logarithms can be partially exponentiated (see \cite{D'Alesio:2014vja} for the quark case):
\begin{align}\label{eq:exponentiation}
\tilde{C}_{g/j}(x,b_T;C_\z^2\m_b^2,\m) \equiv
\exp\le[h_\G(b_T;\m)-h_\g(b_T;\m)\ri]
{\tilde I}_{g/j}(x,b_T;\m)
\,,
\end{align}
where
\begin{align} \label{eq:h}
\frac{d h_\G}{d\ln \m} &= 
\G_{\rm cusp}^A L_T
\,, \quad\quad
\frac{d h_\g}{d\ln \m} = 
\g^{nc} + \G_{\rm cusp}^A \ln C_\z^2
\,.
\end{align}
Choosing $h_{\G(\g)}(b_T;\m_b)=0$, the first few coefficients for the perturbative expansions of $h_{\G(\g)}$ are:
\begin{align}\label{hcoeffs}
h_{\G(\g)}&=\sum_n h_{\G(\g)}^{(n)} \left( \frac{\a_s}{4\pi}\right)^n
\,,
\nn\\
h_\G^{(1)} &=
\frac{1}{4} L_T^2 \G_0^A\,,\quad\quad\quad\quad
h_\G^{(2)} = 
\frac{1}{12} (L_T^3 \G_0^A\b_0+3L_T^2 \G_1^A)\,,\nn\\
h_\G^{(3)}&=
\frac{1}{24}(L_T^4 \G_0^A\b_0^2+2L_T^3 \G_0^A\b_1
+4L_T^3 \G_1^A\b_0+6 L_T^2 \G_2^A)\,,
\nn\\
h_\g^{(1)} &=
\frac{\g_0^{nc}+\G_0^A \ln C_\z^2}{2 \b_0} \le(\b_0 L_T \ri)
\,,
\quad\quad\quad\quad
h_\g^{(2)} =
\frac{\g_0^{nc}+\G_0^A \ln C_\z^2}{4\b_0} \le(\b_0 L_T \ri)^2
+ \le(\frac{\g_1^{nc}+\G_1^A \ln C_\z^2}{2\b_0} \ri) \le(\b_0 L_T \ri)\,,\nn\\
h_\g^{(3)} &=
\frac{\g_0^{nc}+\G_0^A \ln C_\z^2}{6\b_0} \le(\b_0 L_T \ri)^3
+ \frac{1}{2}\le(\frac{\g_1^{nc}+\G_1^A \ln C_\z^2}{\b_0} 
+ \frac{1}{2}\frac{(\g_0^{nc}+\G_0^A \ln C_\z^2)\b_1}{\b_0^2}  \ri)
\le(\b_0 L_T \ri)^2+
\frac{1}{2} \le(\frac{\g_2^{nc}+\G_2^A \ln C_\z^2}{\b_0}\ri) \le(\b_0 L_T \ri)
\,.
\end{align}

After the various steps we have performed, the OPE of gluon TMDPDFs can be re-written as
\begin{align}\label{eq:refactorization}
{\tilde F}_{g/A}^{Pert}(x_A,b_T;\z_A,\m) &=
\exp\le\{
\int_{\m_0}^{\m} \frac{d\bar\m}{\bar\m}
\g_G\le(\as(\bar\m),\ln\frac{\z_A}{\bar\m^2} \ri)\ri\}\,
\le(\frac{\z_A}{C_\z^2\m_b^2}\ri)^{-D_g(b_T;\m_0)}
e^{h_{\G}(b_T;\m_0)-h_{\g}(b_T;\m_0)}
\nn\\
&\times
\sum_{j=q,\bar q, g}
{\tilde I}_{g/j}(x_A,b_T;\m_0) \otimes f_{j/A}(x_A;\m_0)
\,.
\end{align}
The functions $D_g$, $h_\G$ and $h_\g$ above still contain large logarithms $L_T$ that need to be resummed when $\a_s L_T $ is of order 1 (also the coefficients ${\tilde I}_{g/j}$).
We have already calculated the resummed expression for $D_g^R$ in the previous section, and following the same procedure we can derive the resummed expressions for the terms $h_{\G(\g)}$.
Let us consider their evolution equations, Eq.~(\ref{eq:h}):
\begin{align}\label{eq:hR}
h_\G^R(b_T;\m) &=
h_\G(b_T;\m_b)
+ \int_{\m_b}^{\m}\frac{d\bar\m}{\bar\m} \G_{\rm cusp}^AL_T
=
\int_{\a_s(\m_b)}^{\a_s(\mu)}d\a^\prime
\frac{\G_{\rm cusp}^A(\a^\prime)}{\beta(\a^\prime)}
\int_{\a_s(\m_b)}^{\a^\prime}\frac{d\a}{\b(\a)}
\,,\nn \\
h_\g^R(b_T;\m) &=
h_\g(b_T;\m_b)
+ \int_{\m_b}^{\m}\frac{d\bar\m}{\bar\m} (\g^{nc}+\G_{\rm cusp}^A \ln C_\z^2)
=
\int_{\m_b}^{\m}\frac{d\bar\m}{\bar\m} (\g^{nc}+\G_{\rm cusp}^A \ln C_\z^2)
\,.
\end{align}
Notice that we have chosen $h_{\G(\g)}(b_T;\m_b)=0$.
By expanding the $\b$-function and re-writing $\a_s(\m_b)$ in terms of $\a_s(\m)$ at the proper order, as shown in \cite{Echevarria:2012pw} for the quark TMDPDFs case, we can solve these equations and get
\begin{align}\label{eq:hgar}
h_\G^R(b_T;\m) &=
\frac{ \G_0^A (X-(X-1) \ln (1-X))}{2 a_s \beta_0^2}
\nn\\
&
+
\frac{\beta_1 \G_0^A \left(2 X+\ln ^2(1-X)+2 \ln (1-X)\right)-2
\beta_0 \G_1^A (X+\ln (1-X))}{4 \beta_0^3}
\nn\\
&
+ \frac{a_s}{4 \beta_0^4 (1-X)}\left(\beta_0^2 \G_2^A X^2-\beta_0
(\beta_1 \G_1^A (X (X+2)+2 \ln (1-X))
\ri.
\nn\\
&\le.
+\beta_2
\G_0^A ((X-2) X+2 (X-1) \ln (1-X)))
+\beta_1^2 \G_0^A(X+\ln (1-X))^2\right)
\,,
\end{align}
and
\begin{align}\label{eq:hg}
h_\g^R(b_T;\m) &=
-\frac{\g_0^{nc}+\G_0^A\ln C_\z^2}{2\beta_0}\ln(1-X)
+ \frac{1}{2}\le(\frac{a_s}{1-X}\ri) \le[
- \frac{\beta_1(\g_0^{nc}+\G_0^A\ln C_\z^2)}{\beta_0^2} (X+\ln(1-X))
+\frac{\g_1^{nc}+\G_1^A\ln C_\z^2}{\beta_0} X\ri]
\nn\\
&+ \frac{1}{2}
\le(\frac{a_s}{1-X}\ri)^2\le[
\frac{\g_2^{nc}+\G_2^A\ln C_\z^2}{2\beta_0}(X (2-X))
+\frac{\beta_1(\g_1^{nc}+\G_1^A\ln C_\z^2)}{2 \beta_0^2} \le( X (X-2)-2 \ln (1-X)\ri) \ri.
\nn\\
&\le.
+\frac{\beta_2(\g_0^{nc}+\G_0^A\ln C_\z^2)}{2\beta_0^2} X^2
+\frac{\beta_1^2(\g_0^{nc}+\G_0^A\ln C_\z^2)}{2\beta_0^3} (\ln^2(1-X)-X^2)\right]
\,,
\end{align}
where again $a_s=\as/(4\pi)$ and $X=a_s\b_0L_T$.

The resummed expressions we have just found are valid only in the perturbative region of small $b_T$.
With them, we can finally write the resummed TMDPDFs as:
\begin{align}\label{eq:tmdpert}
{\tilde F}_{g/A}^{Pert}(x_A,b_T;\z_A,\m) &=
\exp\le\{
\int_{\m_0}^{\m} \frac{d\bar\m}{\bar\m}
\g_G\le(\as(\bar\m),\ln\frac{\z_A}{\bar\m^2} \ri)\ri\}\,
\le(\frac{\z_A}{C_\z^2\m_b^2}\ri)^{-D_g^R(b_T;\m_0)}
e^{h_{\G}^R(b_T;\m_0)-h_{\g}^R(b_T;\m_0)}
\nn\\
&\times
\sum_{j=q,\bar q, g}
{\tilde I}_{g/j}(x_A,b_T;\m_0) \otimes f_{j/A}(x_A;\m_0)
\,.
\end{align}
Notice that the functions $D_g^R$, $h_\G^R$ and $h_\g^R$ are universal and spin independent, and thus are the same for any of the TMDPDFs in Eq.~\eqref{eq:decompositionips}.
On the contrary, the coefficients ${\tilde I}_{g/j}$ are specific for each TMDPDF, as are the collinear functions $f_{j/A}$ which generate the perturbative tail for each TMDPDF at small $b_T$.

Finally, in order to parametrize the non-perturbative contribution at large $b_T$ we overlap the perturbative expression in Eq.~\eqref{eq:tmdpert} with a non-perturbative model and write
\begin{align}\label{eq:simplemodel}
{\tilde F}_{g/A}(x_A,b_T;\z_A,\m) &= 
{\tilde F}_{g/A}^{Pert}(x_A,b_T;\z_A,\m)\,
{\tilde F}_{}^{NP}(x_A,b_T;\z_A)
\,.
\end{align}
The two functions, perturbative and non-perturbative, extend over the whole impact parameter space.
However, their contributions should dominate in different regions.
The function ${\tilde F}^{NP}$, the non-perturbative model, should be $1$ for $b_T=0$, where the perturbative expression applies, and play an increasingly important role as $b_T$ gets larger.
Moreover, it should be such that it cancels the contribution that comes from the perturbative expression ${\tilde F}_{g/A}^{Pert}$ in the large $b_T$ region, where it does not apply.
In simple terms, there is no problem with extending the perturbative expression to the whole $b_T$-space, since the model is used to correct for it in the non-perturbative region.
This approach was used in~\cite{D'Alesio:2014vja} to perform a global fit of Drell-Yan and $Z$-boson data and extract the unpolarized quark TMDPDFs.

As a final remark, we emphasize the fact that the non-perturbative model in this case is not the same as in the previous subsection.
They parametrize the non-perturbative region in a different way, depending on what is the treatment of the perturbative contribution, and thus they can be different.

\section{Gluon helicity TMDPDF}
\label{sec:tmdhelicity}

The gluon helicity TMDPDF, $g_{1L}^g$, represents the distribution of longitudinally polarized gluons inside a longitudinally polarized hadron.
In Appendix~\ref{app:helicitynlo} we perform for the first time an explicit NLO calculation of this quantity and show that, if defined as in \eq{eq:tmdsdefinition0}, then the rapidity divergences cancel among the collinear and soft matrix elements.
We also perform a NLO calculation on the collinear gluon helicity PDF $g_{j/A}(x;\m)$, which we use to extract the OPE Wilson coefficient of gluon helicity TMDPDF:
\begin{align}\label{eq:opehelicity}
{\tilde g}_{1L}^{g/A}(x_A,b_T;\z_A,\m) &=
\sum_{j=q,\bar q,g} \int_{x_A}^{1}\frac{d\bar x}{\bar x}
{\tilde C}_{g/ j}^{g}(\bar x,b_T;\z_A,\m)\,
g_{j/A}(x_A/\bar x;\m)
+{\cal O}(b_T\lqcd)
\,,
\end{align}
where the longitudinally polarized collinear quark and gluon PDFs are defined as
\begin{align}
g_{q/A}(x;\m) &= 
\frac{1}{2} \int\frac{dy^-}{2\pi}\,
e^{-i\frac{1}{2}y^- xP^+}\,
\sandwich{PS_A}
{\le[{\bar\x}_nW_n\ri](y^-)\frac{\bnslash}{2}\g_5\le[W^\dagger_n\x_n\ri](0)}
{PS_A}
\,,\nn\\
g_{g/A}(x;\m) &= 
\frac{x P^+}{2} (i\e^\perp_{\m\n})\int\frac{dy^-}{2\pi}\,
e^{-i\frac{1}{2}y^-xP^+}\,
\sandwich{PS_A}{\mB_{n\perp}^{\m,a}(y^-) \mB_{n\perp}^{\n,a}(0)}{PS_A}
\,,
\end{align}
with $\x_{n(\bn)}$ the (anti)collinear fermion field.
The result of the coefficient is analogous to the one for the unpolarized gluon TMDPDF, as we will show in next section, apart from the DGLAP splitting kernel (similar result was found in the case of unpolarized and helicity quark TMDPDFs in \cite{Bacchetta:2013pqa}).
It reads
\begin{align}
\label{eq:helcoeffs}
\tilde{C}^g_{g/g} &= 
\frac{\as}{2\pi} \left[ 
C_A\d(1-x)\le( - \frac{1}{2}L_T^2 + L_T\ln\frac{\m^2}{\z} - \frac{\pi^2}{12}\ri)
- L_T\le({\cal P}_{\D g/\D g}-\d(1-x)\frac{\b_0}{2}\ri) -4C_A(1-x) 
\right]
\,,
\nn\\
\tilde{C}^g_{g/q} &= 
\frac{\as}{2\pi} \bigg[
-L_T {\cal P}_{\D g/\D q}
- C_F 2(1-x)
\bigg]
\,,
\end{align}
where the one-loop DGLAP splitting kernels (collected for all polarizations in \cite{Diehl:2013mla}) are
\begin{align}\label{eq:split}
{\cal P}_{\D g/\D g}(x) &=
{\cal P}_{g/g}(x)
- 2 C_A\frac{(1-x)^3}{x}
\,,
\nn\\
{\cal P}_{g/g}(x) &= 
2C_A \le[ \frac{x}{(1-x)_+} + \frac{1-x}{x} + x(1-x) \ri] + \frac{\b_0}{2}\d(1-x) 
\,,\nn\\
{\cal P}_{\D g/\D q}(x) &=
C_F\frac{1-(1-x)^2}{x}
\,.
\end{align}

In order to illustrate the QCD evolution of gluon helicity TMDPDF we choose the resummation scale in impact parameter space.
We also set $\m^2=\z=Q^2$, use the evolution kernel in Eq.~\eqref{eq:evolkernel} and separate the perturbative and non-perturbative contributions in a smooth way as in \eq{eq:nonp-bhat}.
Thus the gluon helicity TMDPDF is given by
\begin{align}
{\tilde g}_{1L}^{g}(x_A,b_T;Q^2,Q) &=
\exp\le\{
\int_{\m_0}^{Q} \frac{d\bar\m}{\bar\m}
\g_G\le(\as(\bar\m),\ln\frac{Q^2}{\bar\m^2} \ri)\ri\}\,
\le(\frac{Q^2}{\z_0}\ri)^{-D_g({\hat b}_T;\m_0)}
\nn\\
&\times
\sum_{j=q,\bar q, g} \int_{x_A}^{1}\frac{d\bar x}{\bar x}
{\tilde C}_{g/j}^{g}(\bar x,{\hat b}_T;\z_0,\m_0)\,
g_{j/A}(x_A/\bar x;\m_0)\,
{\tilde F_{j/A}}^{NP}(x_A,b_T;Q)
\,,
\end{align}
where $\z_0\sim\m_0^2\sim\m_b^2$.
For our numerical studies the ${\hat b}_T$ prescription is
\begin{align}\label{eq:bhatn2}
\hat b_T(b_T) = b_c\le( 1 - e^{-(b_T/b_c)^2} \ri)^{1/2}
\,,\quad\quad
b_c = 1.5~{\rm GeV}^{-1}
\,,
\end{align}
and we implement a simple non-perturbative model
\begin{align}
{\tilde F_{j/A}}^{NP}(x_A,b_T;Q) &=
\exp\le[-b_T^2(\l_g+\l_Q \ln(Q^2/Q_0^2)) \ri]
\,,\quad\quad
Q_0 = 1~{\rm GeV}
\,.
\end{align}
The parameters $\l_g$ and $\l_Q$ have never been extracted from experimental data, and thus we can only guess their values and give predictions by varying them in a reasonable range.
What we know is that $\l_Q$ is the same among all (un)polarized gluon TMDPDFs, because it parametrizes the scale-dependent part of the non-perturbative model, which is related to the large-$b_T$ tail of the universal $D_g$ function.
Notice that for simplicity we have neglected any $x$ dependences in the non-perturbative model.

\begin{table}[h]
\begin{center}
\setlength{\tabcolsep}{0.2cm}
{\renewcommand{\arraystretch}{1.5}
\begin{tabular}{|c||c|c||c|c|c|c||c|c|}
\hline
 & $\G_{\rm cusp}^A$ & $\g^{nc}$ & ${\tilde I}_{g/j}$ &
$D_g^R$ & $h_\G^R$ & $h_\g^R$ &
${\tilde C}_{g/j}$ & $D_g$
\\
\hline\hline
LL &$\as^1$&$\as^0$&$\as^0$&$\as^0$&$\as^{-1}$&$0$&$\as^{0}$&$\as^{0}$
\\
\hline
NLL &$\as^2$& $\as^1$ & $\as^0$&$\as^1$&$\as^0$&$\as^0$&$\as^{0}$&$\as^{1}$
\\
\hline
NNLL&$\as^3$&$\as^2$&$\as^1$&$\as^2$&$\as^1$&$\as^1$&$\as^{1}$&$\as^{2}$
\\
\hline
\end{tabular}
}
\caption{Perturbative orders in logarithmic resummations, both for for resummations in momentum space and in impact parameter space.
\label{tab:resummation}}
\end{center}
\end{table}

\begin{figure}[t]
\begin{center}
\includegraphics[width=0.35\textwidth]{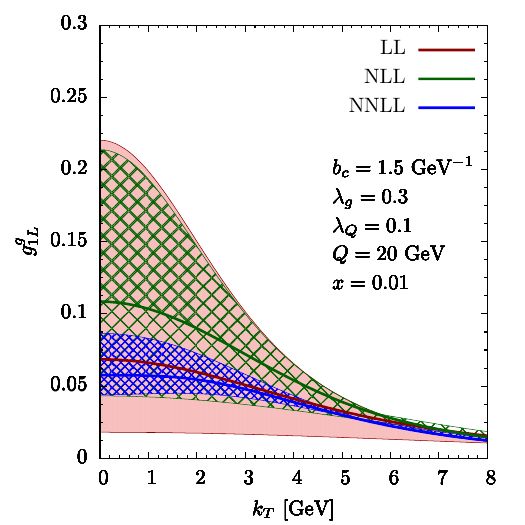}
\end{center}
\caption{\it
The gluon helicity TMDPDF $g_{1L}^g$ at $Q=20$ GeV, $x=0.01$ and with the non-perturbative parameters chosen to be $\l_g= 0.3$ and $\l_Q= 0.1$.
The bands come from varying independently both the resummation scale $\m_0$ and the rapidity scale $\z_0$ by a factor of $2$ around their default value, and taking the maximum variation.
}
\label{fig:helicity_evol}
\end{figure}

The gluon helicity TMDPDF is shown in Fig.~\ref{fig:helicity_evol} at
$x = 0.01$, $Q=20$ GeV, with the non-perturbative parameters
$b_c = 1.5$ GeV$^{-1}$, $\lambda_g = 0.3$, $\lambda_Q= 0.1$.
We use the latest available parametrizations of collinear gluon and quark helicity PDFs~\cite{deFlorian:2014yva} at NLO for our numerical analysis.
The running of the strong coupling is implemented at NNNLO with the MSTW routine~\cite{Martin:2009iq}, with a variable flavor number scheme with $m_c=1.4$ GeV and $m_b=4.75$ GeV.
The input value for the strong coupling is set to $\as(M_Z)=0.1185$, and we impose a lower cutoff for the running scale $\mu$ such that it never goes below 1 GeV.
The bands come from varying both the rapidity and the resummation scales by a factor of $2$ around their default value, and keeping the largest variation for each point in $k_T$.
It is clear that the theoretical uncertainty gets reduced as we increase the resummation accuracy by including more perturbative ingredients, as schematically illustrated in Table~\ref{tab:resummation}.

\begin{figure}[t]
\begin{center}
\includegraphics[width=0.35\textwidth]{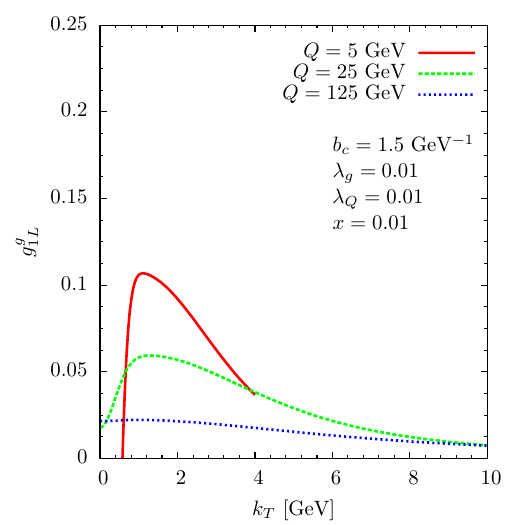}
\quad
\includegraphics[width=0.35\textwidth]{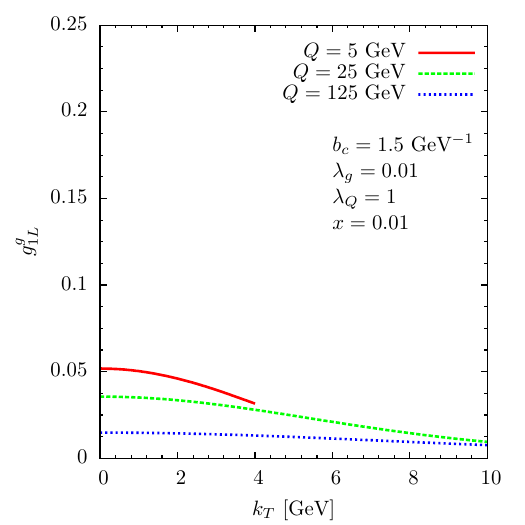}
\\
\includegraphics[width=0.35\textwidth]{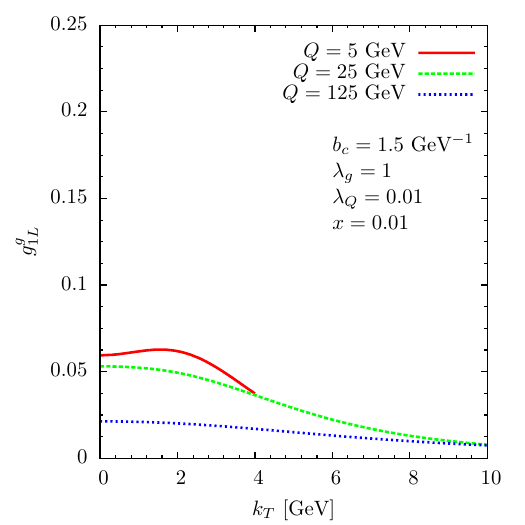}
\quad
\includegraphics[width=0.35\textwidth]{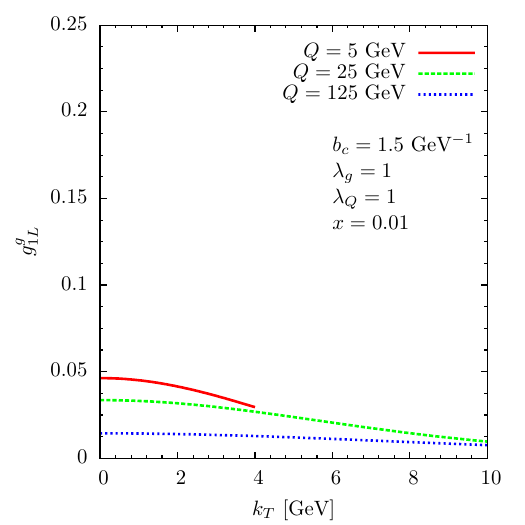}
\end{center}
\caption{\it
The gluon helicity TMDPDF $g_{1L}^g$ at $x=0.01$ for different values of the evolution scale and the non-perturbative parameters, at $\sqrt{s}=8$ TeV.
All curves are given at NNLL accuracy.
}
\label{fig:helicity_np}
\end{figure}

In Fig.~\ref{fig:helicity_np} we show the gluon helicity TMDPDF at $x=0.01$,
for different values of the energy scale and the non-perturbative parameters,
at NNLL accuracy. 
In order to be consistent with the factorization theorem we have cut the curves so that the condition $k_T\ll Q$ is fulfilled. 
As can be seen, the larger the scale, the wider is the distribution.
Moreover, the larger the value of the non-perturbative parameters, the smaller the helicity distribution is at low transverse momentum.
It is interesting to notice that choosing small and equal non-perturbative parameters, gives a qualitatively different gluon helicity TMDPDF at low scales, as in the upper-left panel.
This is due to the fact that with these parameters the non-perturbative input is very small, and thus the helicity TMDPDF is given mostly by its perturbative expression. 
However, at low scales the main contribution to the helicity distribution originates from a region where the impact parameter is large and the perturbative result alone should not be trusted.

These results have been obtained using the OPE coefficients that have been calculated for the first time in the present work.
They are an important perturbative ingredient that will allow us to better fix the non-perturbative parameters with new measurements, that could be performed at RHIC or at the future AFTER@LHC or EIC with longitudinally polarized hadron beams.

\section{Gluon TMDPDFs in an Unpolarized Hadron}
\label{sec:tmdsunpolhadron}

As shown in Eq.~\eqref{eq:decompositionips} there are two gluon distributions that contribute at leading-twist in the case of an unpolarized hadron: the unpolarized ($f_1^g$) and the linearly polarized ($h_{1}^{\perp g}$) ones.
The latter was introduced in \cite{Mulders:2000sh} and implemented for the first time in the resummation of gluon-gluon fusion process in impact parameter space in \cite{Catani:2010pd,Catani:2011kr}. 
In Appendices~\ref{app:tmdpdfnlo} and~\ref{app:linearnlo} we perform an explicit NLO calculation of those distributions using their proper definition in \eq{eq:tmdsdefinition}, and we show that they are free from rapidity divergences when the collinear and soft matrix elements are combined properly.

As described in the previous section for the gluon helicity TMDPDF, both $f_1^g$ and $h_{1}^{\perp g}$ in the small $b_T$ region can be factorized in terms of collinear functions, which in this case are just the unpolarized collinear gluon/quark PDFs:
\begin{align}\label{eq:opeunpol}
{\tilde f}_{1}^{g/A}(x_A,b_T;\z_A,\m) &=
\sum_{j=q,\bar q, g} \int_{x_A}^{1}\frac{d\bar x}{\bar x}
{\tilde C}_{g/ j}^{f}(\bar x,b_T;\z_A,\m)\,
f_{j/A}(x_A/\bar x;\m)
+{\cal O}(b_T\lqcd)
\,,\nn\\
{\tilde h}_{1}^{\perp g/A\,(2)}(x_A,b_T;\z_A,\m) &=
\sum_{j=q,\bar q, g} \int_{x_A}^{1}\frac{d\bar x}{\bar x}
{\tilde C}_{g/ j}^{h}(\bar x,b_T;\z_A,\m)\,
f_{j/A}(x_A/\bar x;\m)
+{\cal O}(b_T\lqcd)
\,,
\end{align}
where the unpolarized collinear PDFs are defined as
\begin{align}
f_{q/A}(x;\m) &= 
\frac{1}{2} \int\frac{dy^-}{2\pi}\,
e^{-i\frac{1}{2}y^- xP^+}\,
\sandwich{PS_A}{\le[{\bar\x}_nW_n\ri](y^-)\frac{\bnslash}{2}\le[W^\dagger_n\x_n\ri](0)}{PS_A}
\,,\nn\\
f_{g/A}(x;\m) &= 
\frac{x P^+}{2} \int\frac{dy^-}{2\pi}\,
e^{-i\frac{1}{2}y^-xP^+}\,
\sandwich{PS_A}{\mB_{n\perp}^{\m,a}(y^-) \mB_{n\perp\m}^{a}(0)}{PS_A}
\,.
\end{align}
Note that the TMDPDFs for the unpolarized gluon and the linearly polarized gluon are both matched onto the same PDF, but the first non-zero order of the matching coefficient for the linearly polarized gluon is one order higher in $\alpha_s$ than for the unpolarized gluon.
In Appendices~\ref{app:tmdpdfnlo} and~\ref{app:linearnlo} we obtain their matching coefficients at NLO by subtracting the collinear PDFs at the same order.
Moreover, in Section~\ref{sec:refactorization} we have shown that the OPE coefficients for TMDs can be further refactorized, and thus the previous OPEs can be written, setting $\m^2=\z=Q^2$ and using the evolution kernel in Eq.~\eqref{eq:evolkernel}, as
\begin{align}\label{eq:tmdspert}
{\tilde f}_{1}^{g/A}(x_A,b_T;Q^2,Q) &=
\exp\le\{
\int_{\m_0}^{Q} \frac{d\bar\m}{\bar\m}
\g_G\le(\as(\bar\m),\ln\frac{Q^2}{\bar\m^2} \ri)\ri\}\,
\le(\frac{Q^2}{\z_0}\ri)^{-D_g(b_T;\m_0)}
\nn\\
&\times
\sum_{j=q,\bar q, g} \int_{x_A}^{1}\frac{d\bar x}{\bar x}
{\tilde C}_{g/ j}^{f}(\bar x,b_T;\z_0,\m_0)\,
f_{j/A}(x_A/\bar x;\m_0)
+{\cal O}(b_T\lqcd)
\,,\nn\\
{\tilde h}_{1}^{\perp g/A\,(2)}(x_A,b_T;Q^2,Q) &=
\exp\le\{
\int_{\m_0}^{Q} \frac{d\bar\m}{\bar\m}
\g_G\le(\as(\bar\m),\ln\frac{Q^2}{\bar\m^2} \ri)\ri\}\,
\le(\frac{Q^2 b_T^2}{4e^{-2\g_E}}\ri)^{-D_g(b_T;\m_0)}
\nn\\
&\times
\sum_{j=q,\bar q, g} \int_{x_A}^{1}\frac{d\bar x}{\bar x}
{\tilde C}_{g/ j}^{h}(\bar x,b_T;\z_0,\m_0)\,
f_{j/A}(x_A/\bar x;\m_0)
+{\cal O}(b_T\lqcd)
\,.
\end{align}
The perturbative coefficients ${\tilde C}_{g/ j}^{f,h}$ are given in Appendices~\ref{app:tmdpdfnlo} and~\ref{app:linearnlo}, the one-loop DGLAP splitting kernel ${\cal P}_{g/g}$ is given in \eq{eq:split} and 
\begin{align}
{\cal P}_{g/q}(x) &= 
C_F\frac{1+(1-x)^2}{x}
\,.
\end{align}

\begin{figure}[t]
\begin{center}
\includegraphics[width=0.35\textwidth]{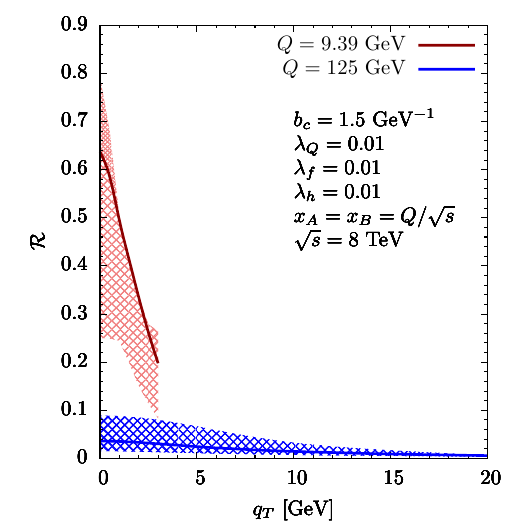}
\quad
\includegraphics[width=0.35\textwidth]{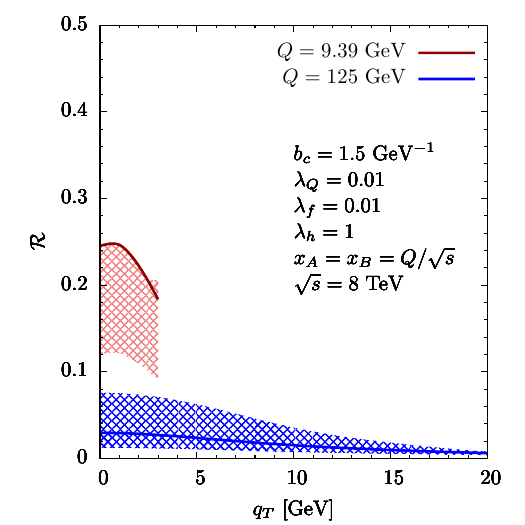}
\\
\includegraphics[width=0.35\textwidth]{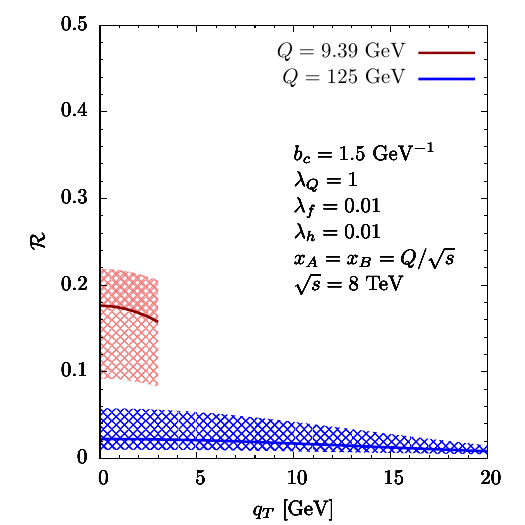}
\quad
\includegraphics[width=0.35\textwidth]{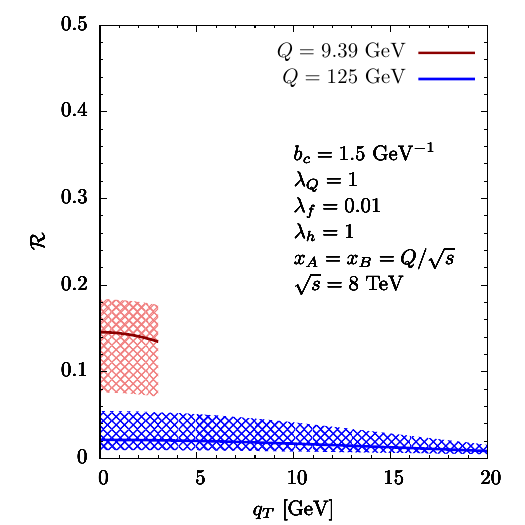}
\end{center}
\caption{\it
Ratio ${\cal R}$ for different values of the non-perturbative parameters $\l_{f(h)}$ and $\l_Q$, at the relevant scales for Higgs boson and $\eta_{b}$ production. 
The curves are calculated at NNLL accuracy and for $\sqrt{s}=8$ TeV.
The bands come from varying independently both the resummation scale $\m_0$ and the rapidity scale $\z_0$ by a factor of $2$ around their default value, and taking the maximum variation.
Notice the differences in scales on the vertical axes.
}
\label{fig:ratio}
\end{figure}

The contribution of unpolarized and/or linearly polarized gluon distributions in unpolarized hadron-hadron collisions depends on the process under study and has been discussed in several works~\cite{Boer:2011kf,Boer:2013fca,Pisano:2013cya,Dunnen:2014eta,Boer:2014tka}.
In this work we focus on the production of Higgs boson and $C$-even pseudoscalar bottonium state $\eta_b$~\cite{Ma:2012hh}, since for the production of $P$-wave quarkonium states (like $\chi_{b0}$) there are arguments that suggest a breaking of the factorization~\cite{Ma:2014oha}.
In the considered cases, Higgs boson and $\eta_b$ production, both unpolarized and linearly polarized distributions play a role, and thus one can investigate their relative contribution to the cross-section.
We use our results to quantify the contribution of linearly polarized gluons, considering the following ratio~\footnote{The moments of TMDPDFs, and in particular the second moment of $h_{1}^{\perp g}$, are defined in \eq{eq:moments}.}:
\begin{align}
{\cal R}(x_A,x_B,q_T;Q) &=
\frac{\int d^2\vecb b_T\,e^{-i\vecbe q_T\cd \vecbe b_T}
{\tilde h}_{1}^{\perp g/A(2)}(x_A,b_T;Q^2,Q)\,
{\tilde h}_{1}^{\perp g/B(2)}(x_B,b_T;Q^2,Q)}
{\int d^2\vecb b_T\,e^{-i\vecbe q_T\cd \vecbe b_T}
{\tilde f}_{1}^{g/A}(x_A,b_T;Q^2,Q)\,{\tilde f}_{1}^{g/B}(x_B,b_T;Q^2,Q)}
\,,
\end{align}
where the numerator and denominator are the two terms in the factorized cross section which determine the relative contribution from linearly polarized and unpolarized gluons to the cross section, for both Higgs boson and $C$-even pseudoscalar bottonium production.
In order to compute this quantity we will insert the TMDs as in \eq{eq:tmdspert}, choosing $\z_0\sim\m_0^2\sim\mu_b^2$ and using the ${\hat b}_T$ prescription to separate the perturbative from non-perturbative contributions as in \eq{eq:bhatn2}.
The latter will be parametrized as:
\begin{align}
{\tilde F_{j/A}}^{f,NP}(x_A,b_T;Q) &=
\exp\le[-b_T^2(\l_f+\l_Q \ln(Q^2/Q_0^2)) \ri]
\,,\quad\quad
Q_0 = 1~{\rm GeV}\,,
\nn\\
{\tilde F_{j/A}}^{h,NP}(x_A,b_T;Q) &=
\exp\le[-b_T^2(\l_h+\l_Q \ln(Q^2/Q_0^2)) \ri]
\,,\quad\quad
Q_0 = 1~{\rm GeV}
\,,
\end{align}
similar to the model used previously for the helicity TMDPDF.
Notice that the parameter $\l_Q$ is the same for both functions, since the evolution is universal among all (un)polarized TMDPDFs, that is, their scale-dependence is the same.
Having precise estimates for this ratio will help us predict the measurability of both unpolarized and linearly polarized gluon distributions in a given process (or scale), which is the final goal.
Using our resummation scheme and the perturbative ingredients at the highest possible order we provide accurate predictions for this quantity.

In Fig.~\ref{fig:ratio} we show our results for the ratio ${\cal R}$ at the relevant scales for the transverse momentum distributions of Higgs boson and $\eta_{b}$, all at NNLL accuracy.
We used the MSTW08nnlo set~\cite{Martin:2009iq} and selected different values for the non-perturbative parameters in order to check their impact on the result.
The running of the strong coupling is implemented at NNNLO with the MSTW routine, with a variable flavor number scheme with $m_c=1.4$ GeV and $m_b=4.75$ GeV.
The input value for the strong coupling is set to $\as(M_Z)=0.1185$, and we impose a lower cutoff for the running scale $\mu$ such that it never goes below 0.4 GeV.
Comparing our results to the ones presented in \cite{Boer:2014tka}, we have included the contribution of quark PDFs to the collinear expansion of gluon TMDPDFs (through ${\tilde C}^{f,h}_{g/q}$ in \eq{eq:tmdspert}) and also higher order perturbative ingredients, performing the resummation consistently at NNLL accuracy. 
The uncertainty bands also represent an improvement with respect to the results in \cite{Boer:2014tka}: they allow us to better quantify what is the effect of non-perturbative contributions relative to the scale uncertainty, and whether experimental data can be used to determine the non-perturbative parameters or distinguish between different models of the non-perturbative input.
The bands are obtained by independently varying the scales $\z_0$ and $\m_0$ around their default value by a factor of 2, and plotting the maximum uncertainty for each point in $q_T$.

In order to estimate the impact on the ratio of the different non-perturbative parameters, we have chosen several values in a sensible range and selected some combinations in limiting cases.
First, the parameters should be positive, since the gluon distributions are supposed to vanish at large $b_T$.
Second, given the values found for similar models in the case of quark TMDPDFs (see, e.g.,~\cite{Konychev:2005iy}), we have chosen a maximum value of 1.
Finally, since our goal is to estimate the contribution of linearly polarized gluons as compared to unpolarized ones, we have chosen the following limiting cases:
\begin{itemize}
\item[(i)] 
$\l_Q=0.01\,,\,\,\l_f=\l_h=0.01$. Small evolution parameter $\l_Q$ and similar and small parameters $\l_f$ and $\l_h$.
\item[(ii)]
$\l_Q=0.01\,,\,\,\l_f=0.01\,,\,\,\l_h=1$. Small evolution parameter $\l_Q$ and $\l_h\gg\l_f$.
\item[(iii)] 
$\l_Q=1\,,\,\,\l_f=\l_h=0.01$. Large evolution parameter $\l_Q$ and similar and small parameters $\l_f$ and $\l_h$.
\item[(iv)]
$\l_Q=1\,,\,\,\l_f=0.01\,,\,\,\l_h=1$. Large evolution parameter $\l_Q$ and $\l_h\gg\l_f$.
\end{itemize}

The outcome of the numerical study is clear: the lower the scale the more contribution we have from linearly polarized gluons, although this contribution depends on the value of the non-perturbative parameters, which will have to be fixed by fitting experimental data.
At the Higgs boson scale the effect of linearly polarized gluons is small, around 1-9\%, making it harder to extract their non-perturbative parameters from experimental data.
At lower scales, as in the production of $\eta_{b}$, their role is enhanced, from 10\% up to 70\%, and thus experimental data can better determine them.
However, it seems plausible that their non-perturbative parameters could be fixed in the near future by properly combining experimental data for different experiments and at different scales.
Thus the framework introduced in this paper, with the proper definition of gluon TMDPDFs and their QCD evolution, will be crucial in order to consistently address different processes in terms of the same hadronic quantities and properly extract their non-perturbative parameters.

\section{Higgs boson $q_T$-distribution}
\label{sec:pheno}

After analyzing the contribution of linearly polarized gluons for $\eta_b$ and Higgs boson production in unpolarized hadron-hadron collisions, we apply our results to provide some predictions for the Higgs boson transverse momentum distribution at the LHC.
The cross-section for this process can be easily obtained from \eq{eq:factth} if we consider unpolarized protons:
\begin{align}\label{eq:higgsdistribution}
\frac{d\sigma}{dy\,d^2q_\perp} &= 
2\s_0(\mu)\,C_t^2(m_t^2,\mu) H(m_H^2,\m)\,
\frac{1}{(2\pi)^2} \int d^2y_\perp\,e^{i\vecbe q_\perp\cd\vecbe y_\perp} 
\nn\\
&
\times
\frac{1}{2}\le[
{\tilde f}_1^{g/A}(x_A,b_T;\z_A,\m)\,
{\tilde f}_1^{g/B}(x_B,b_T;\z_B,\m)
+
{\tilde h}_1^{\perp\, g/A (2)}(x_A,b_T;\z_A,\m)\,
{\tilde h}_1^{\perp\, g/B (2)}(x_B,b_T;\z_B,\m)
\ri]
+
{\cal O}(q_T/m_H)
\,.
\end{align}

It is well-known that the evolution kernel suppresses the TMDPDFs at large $b_T$, and that this effect is enhanced the larger the relevant hard scale $Q$ is, in this case $m_H$ \cite{Echevarria:2012pw} (see also, e.g., the discussion in \cite{Qiu:2000hf} in the context of the Collins-Soper-Sterman approach).
Therefore, the larger the $Q$ the more insensitive is the resummed expression to non-perturbative contributions at large $b_T$.
Based on this, we fix the resummation scale in momentum space, $\m_0=Q_0+q_T$ (with $Q_0=2~\gev$), and write
\begin{align}\label{eq:tmdspert2}
{\tilde f}_{1}^{g/A}(x_A,b_T;Q^2,Q) &=
\exp\le\{
\int_{\m_0}^{Q} \frac{d\bar\m}{\bar\m}
\g_G\le(\as(\bar\m),\ln\frac{Q^2}{\bar\m^2} \ri)\ri\}\,
\le(\frac{Q^2}{\z_0}\ri)^{-D_g^R(b_T;\m_0)}
e^{h_\G^R(b_T;\m_0)-h_\g^R(b_T;\m_0)}
\nn\\
&\times
\sum_{j=q,\bar q, g} \int_{x_A}^{1}\frac{d\bar x}{\bar x}
{\tilde I}_{g/ j}^{f}(\bar x,b_T;\m_0)\,
f_{j/A}(x_A/\bar x;\m_0)
+{\cal O}(b_T\lqcd)
\,,\nn\\
{\tilde h}_{1}^{\perp g/A\,(2)}(x_A,b_T;Q^2,Q) &=
\exp\le\{
\int_{\m_0}^{Q} \frac{d\bar\m}{\bar\m}
\g_G\le(\as(\bar\m),\ln\frac{Q^2}{\bar\m^2} \ri)\ri\}\,
\le(\frac{Q^2}{\z_0}\ri)^{-D_g^R(b_T;\m_0)}
e^{h_\G^R(b_T;\m_0)-h_\g^R(b_T;\m_0)}
\nn\\
&\times
\sum_{j=q,\bar q, g} \int_{x_A}^{1}\frac{d\bar x}{\bar x}
{\tilde I}_{g/ j}^{h}(\bar x,b_T;\m_0)\,
f_{j/A}(x_A/\bar x;\m_0)
+{\cal O}(b_T\lqcd)
\,,
\end{align}
where $\z_0\sim \m_b^2$.
The perturbative coefficients ${\tilde I}_{f(h)}$ are derived from the results in Appendices~\ref{app:tmdpdfnlo} and~\ref{app:linearnlo}:
\begin{align}\label{eq:coeffsfh}
{\tilde I}_{g/g}^{f}(x,b_T;\m) &=
\delta(1-x)
+ \frac{\as}{2\pi} \left[
-{\cal P}_{g/g} L_T - C_A\frac{\pi^2}{12}\d(1-x) \right]
\,,
\nn\\
{\tilde I}_{g/q}^{f}(x,b_T;\m) &=
\frac{\as}{2\pi}\le[ - {\cal P}_{g/q} L_T + C_F x\right]
\,,
\nn\\
{\tilde I}_{g/g}^{h}(x,b_T;\m) &=
-\frac{\as}{2\pi} \left[
2C_A\,\frac{1-x}{x} \right]
\,,
\nn\\
{\tilde I}_{g/q}^{h}(x,b_T;\m) &=
-\frac{\as}{2\pi}\le[
2C_F\,\frac{1-x}{x} \right]
\,.
\end{align}

We parametrize the TMDPDFs as in \eq{eq:simplemodel}, which allows us to exploit the perturbative results without using any prescription, like the ${\hat b}_T$.
This procedure was already used in \cite{D'Alesio:2014vja} to perform a global fit of Drell-Yan data, to obtain the non-perturbative parameters of unpolarized quark TMDPDFs. 
Following the same procedure, and leaving some room for small non-perturbative effects, we multiply the TMDPDFs in \eq{eq:tmdspert2} by the following non-perturbative models:
\begin{align}\label{eq:models}
{\tilde F}^{\rm NP}_f(x,b_T;Q) &=
e^{-\b_f b_T}
\,,
\nn\\
{\tilde F}^{\rm NP}_h(x,b_T;Q) &=
e^{-\b_h b_T}
\,,
\end{align}
where we have neglected the dependence on $x$ and $Q$ for simplicity.

The resummation of large logarithms in the cross-section in \eq{eq:higgsdistribution} is done by evaluating each perturbative coefficient at its natural scale and then evolving them up to a common scale by using their relevant anomalous dimensions.
For the TMDPDFs the resummation was already discussed before and led to \eq{eq:tmdspert2}.
The natural scale for the coefficient $C_t$ is $\m_t \sim m_t$, and its evolution is presented in Appendix~\ref{app:ads}.
For $C_H(-q^2,\m)$ (remember that $H=|C_H|^2$) it was discussed in \cite{Ahrens:2008qu} that the choice $\m_H^2\sim -m_H^2$ leads to a better convergence of the resummed expression, and thus we apply that procedure in our numerical study, using the anomalous dimensions that appear in Appendix~\ref{app:ads}.
At NLL accuracy we take the coefficients at LO, and at NNLL accuracy at NLO, combined with the TMDPDFs at the corresponding order given in Table~\ref{tab:resummation}.

\begin{figure}[t]
\begin{center}
\includegraphics[width=0.35\textwidth]{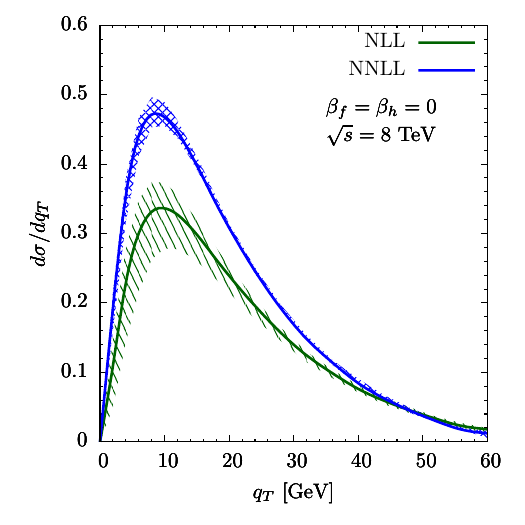}
\quad
\includegraphics[width=0.35\textwidth]{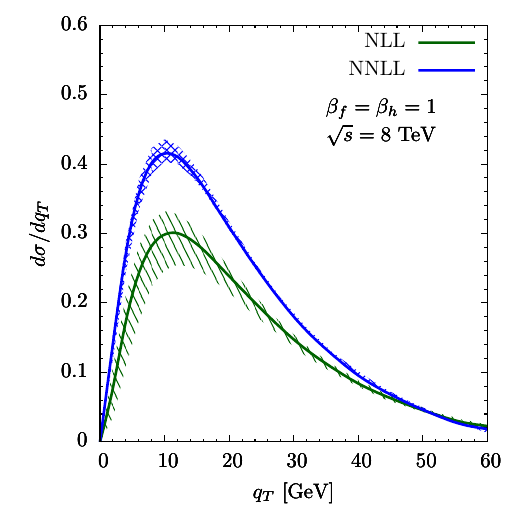}
\end{center}
\caption{\it
Cross-section for different values of the non-perturbative parameters $\b_{f,h}$ and the resummation done in momentum space, with $\m_0\sim q_T$. 
We have $\sqrt{s}=8~{\rm TeV}$ and $m_H=125~{\rm GeV}$.
The bands come from varying independently both the resummation scale $\m_0$ and the rapidity scale $\z_0$ by a factor of $2$ around their default value, and taking the maximum variation.
}
\label{fig:xsection}
\end{figure}

In Figure~\ref{fig:xsection} we show the Higgs boson transverse momentum distribution at $\sqrt{s}=8$ TeV, for different values of the non-perturbative parameters and both at NLL and NNLL accuracies.
We have chosen $\b_h=\b_f$ for simplicity, given that in the previous section we showed that the impact of linearly polarized gluons at the Higgs boson scale is small.
The choices $\b_f=\b_h=0$ and $\b_f=\b_h=1$ give the most extreme scenarios, where the total non-perturbative contribution of both functions is zero or large.
We used the MSTW08nnlo set~\cite{Martin:2009iq} for the input PDFs.
The running of the strong coupling is implemented at NNNLO with the MSTW routine, with a variable flavor number scheme with $m_c=1.4$ GeV and $m_b=4.75$ GeV.
The input value for the strong coupling is set to $\as(M_Z)=0.1185$, and we impose a lower cutoff for the running scale $\mu$ such that it never goes below $Q_0=2$ GeV.
The bands come from varying both the resummation scale $\m_0$ and the rapidity scale $\z_0$ by a factor of 2 around their default values, which gives a much larger contribution than the variation of the scales $\m_t$ and $\m_H$.  
The bands at NNLL get smaller than the ones at NLL, but there is no overlap between them.
This is because we have exponentiated the rapidity scale dependence of the parameter $C_\z$ through the resummed $h_\g^R$ in \eq{eq:hg}.
This exponentiation makes the cross-section at a given resummation order contain some contributions of higher orders, and thus the NLL band, in particular, is smaller.
At NNLL this issue is much less relevant, as will be clear below when comparing with the prediction with the resummation in impact parameter space.

If we compare the two panels in Figure~\ref{fig:xsection}, we see that the impact of the non-perturbative contribution leads to a significant change of the distribution.
However the choice $\beta_{f,h}=1$ is rather extreme, since the non-perturbative model in \eq{eq:models}, which is an exponential function, induces rather large corrections to the perturbative expression in the low $b_T$ region, exactly where one would expect it to work better.
Thus, given that the non-perturbative parameters should probably be smaller, we conclude that the Higgs boson transverse momentum distribution is not very sensitive to those parameters. 
The same conclusion was drawn in \cite{Becher:2012yn}, where a Gaussian model was used to parametrize the non-perturbative contributions, and which led to an almost negligible impact on the distribution.
This is easy to understand, since the Gaussian function in the low $b_T$ region is closer to 1 than the exponential function and therefore has an even smaller impact.

\begin{figure}[t]
\begin{center}
\includegraphics[width=0.35\textwidth]{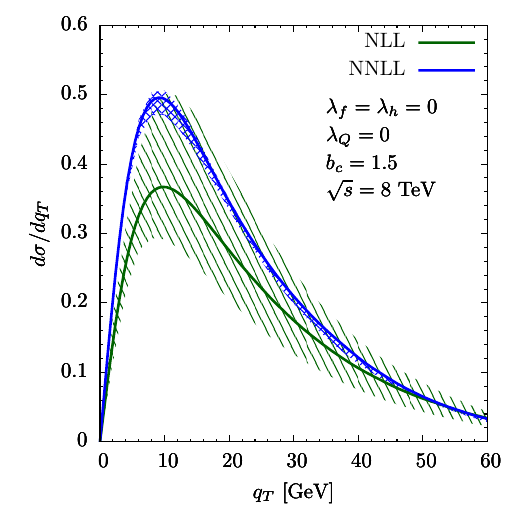}
\quad
\includegraphics[width=0.35\textwidth]{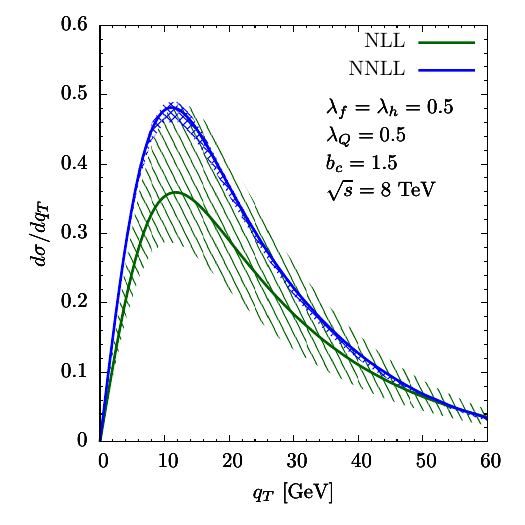}
\end{center}
\caption{\it
Cross-section for different values of the non-perturbative parameters $\l_{f,h,Q}$ and the resummation done in impact parameter space, with $\m_0\sim \m_b$. 
We have $\sqrt{s}=8~{\rm TeV}$ and $m_H=125~{\rm GeV}$.
The bands come from varying independently both the resummation scale $\mu_0$ and the rapidity scale $\z_0$ by a factor of $2$ around their default value, and taking the maximum variation.
}
\label{fig:xsection2}
\end{figure}

Let us now turn our attention to Figure~\ref{fig:xsection2}, where we present a similar prediction to the one already discussed, but with the resummation performed in impact parameter space.
We use here the same settings for the PDFs and the running of the strong coupling as in Figure~\ref{fig:xsection}.
The relevant resummed expressions for the unpolarized and linearly polarized gluon distributions were given in \eq{eq:tmdspert}.
As can be seen, the bands at NLL are now bigger compared to the previous approach, and overlap with the NNLL bands.
Again, comparing the two panels in this figure we see that the effect of the explored non-perturbative parameters is rather small.
Notice that the NNLL curves within the two approaches, in Figures~\ref{fig:xsection} and~\ref{fig:xsection2}, are compatible, and consistent with the recent results found in \cite{Neill:2015roa}.

\begin{figure}[t]
\begin{center}
\includegraphics[width=0.35\textwidth]{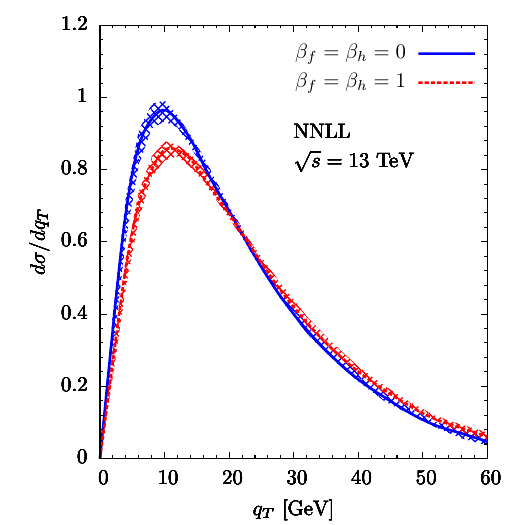}
\quad
\includegraphics[width=0.35\textwidth]{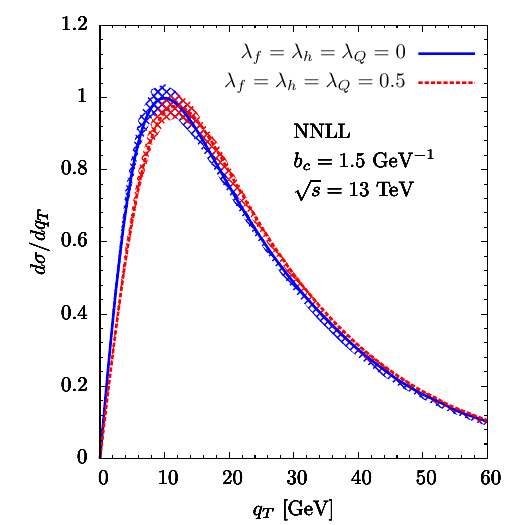}
\end{center}
\caption{\it
Cross-section at $\sqrt{s}=13~{\rm TeV}$ and NNLL accuracy for different values of the non-perturbative parameters $\b_{f,h}$ and $\l_{f,h,Q}$, depending on the resummation scheme used. 
On the left plot the resummation is done in momentum space, while on the right plot it is done in impact parameter space.
The bands come from varying independently both the resummation scale $\m_0$ and the rapidity scale $\z_0$ by a factor of $2$ around their default value, and taking the maximum variation.
}
\label{fig:xsection13}
\end{figure}

Finally, in Figure~\ref{fig:xsection13} we show the predictions for the distribution at $\sqrt{s}=13$ TeV, at NNLL accuracy and again for extreme values of the non-perturbative parameters, with both resummation approaches.
The cross-section is bigger than at $\sqrt{s}=8$ TeV, but the same conclusions regarding the sensitivity to the non-perturbative parameters apply:
the range in parameter variation shown in the left figure is rather large, and it seems unlikely that experimental measurements of the Higgs $q_T$ distribution at the LHC will be precise enough to fix the non-perturbative parameters of gluon TMDPDFs, apart from excluding the most vivid parameter values.

\section{Conclusions}
\label{sec:conclusions}

Using the effective field theory methodology we have derived the factorization theorem for the Higgs boson transverse momentum distribution in hadron-hadron collisions with general polarizations, $A(P,S_A)+B(\bP,S_B)\to H(m_H,q_T)+X$.
By doing so, we have provided the proper definition of all the leading-twist (un)polarized gluon TMDPDFs which could contribute, adequately combining the relevant collinear and (part of) soft matrix elements in order to cancel the spurious rapidity divergences.
We have explicitly shown at NLO that, as expected, those rapidity divergences cancel in the proper definition of gluon TMDPDFs for three different distributions: unpolarized gluons in an unpolarized hadron ($f_1^g$), linearly polarized gluons in an unpolarized hadron ($h_1^{\perp g}$) and longitudinally polarized gluons in a longitudinally polarized hadron ($g_{1L}^g$).
Having at our disposal the proper definition of gluon TMDPDFs is crucial in order to consistently analyse different processes where they appear.

From the structure of the factorization theorem derived, we conclude that the evolution of all leading-twist (un)polarized gluon TMDPDFs is universal, i.e., the same evolution kernel can be applied to evolve any of them.
Moreover, given the currently known perturbative ingredients we have performed the resummation of large logarithms contained in this evolution kernel up to NNLL accuracy.

TMDPDFs are functions that contain perturbatively calculable information when the transverse momentum is in the perturbative domain.
In this work we have considered all gluon TMDPDFs and discussed their operator product expansion in terms of collinear functions.
The OPE Wilson coefficients depend on the particular distribution but we have shown that part of them is the same for all TMDPDFs.
We have furthermore resummed those universal pieces at NNLL accuracy, increasing our control over the perturbative ingredients of the TMDPDFs.
Moreover we have derived, for the first time, the NLO Wilson coefficient for the gluon helicity TMDPDF $g_{1L}^g$, which will allow more accurate phenomenological studies of this quantity in the future, e.g., at RHIC, AFTER@LHC or EIC.
We have also derived the OPE Wilson coefficients for $f_1^g$ and $h_1^{\perp g}$ in the framework presented in this paper.

Using the obtained results we have performed a numerical study of the contribution of linearly polarized gluons for the productions of $\eta_b$ and Higgs boson in unpolarized hadron-hadron collisions.
The major conclusion is that the larger the relevant hard scale is, the less sensitive is the observable to their non-perturbative contribution, and therefor harder to extract.
Thus one would need to combine low- and high-energy experimental data and properly implement the QCD evolution of gluon TMDPDFs in order to extract it.
On the other hand, the fact that at large scales the transverse momentum distributions are less sensitive to the non-perturbative parameters of the TMDPDFs allows us to obtain accurate predictions even if currently there is no information on these parameters.

Finally we have provided some predictions for the Higgs boson transverse momentum distribution at the LHC, both at $\sqrt{s}=8$ TeV and $\sqrt{s}=13$ TeV, using the formalism presented in this paper, i.e., expressing it in terms of well-defined gluon TMDPDFs.
We have studied the impact of non-perturbative contributions on the distribution and have shown that the sensitivity to them is very small.

\section*{Acknowledgements}

We thank Daniel Boer and Wouter Waalewijn for useful discussions, as well as Maarten Buffing, Markus Diehl, Daniel Gutierrez, Ignazio Scimemi and Alexey Vladimirov.
We are also grateful to Daniel de Florian, Rodolfo Sassot, Marco Stratmann and Werner Vogelsang for providing us with their code.
We acknowledge financial support from the European Community under the FP7 ``Ideas'' program QWORK (contract 320389). 
M.G.E. is supported by the ``Stichting voor Fundamenteel Onderzoek der Materie'' (FOM), which is financially supported by the ``Nederlandse Organisatie voor Wetenschappelijk Onderzoek'' (NWO). 
C.P. acknowledges support by the ``Fonds Wetenschappelijk Onderzoek - Vlaanderen'' (FWO) through a postdoctoral Pegasus Marie Curie Fellowship.
Figures were made using JaxoDraw \cite{Binosi:2003yf}, parts of the calculation were preformed with the aid of FeynCalc \cite{Mertig:1990an}.

\appendix

\section{OPE of $f_1^g$ at NLO}
\label{app:tmdpdfnlo}

\begin{figure}[t]
\begin{center}
\includegraphics[width=\textwidth]{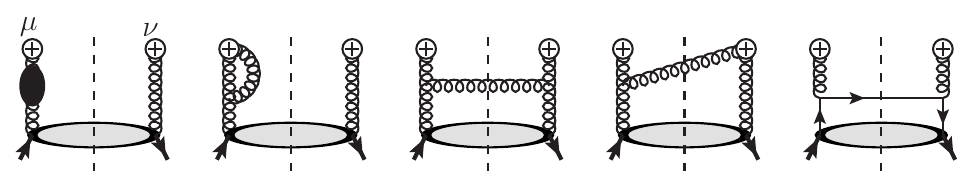}
\\
(a)\hspace{3.2cm}(b)\hspace{3.2cm}(c)\hspace{3.2cm}(d)\hspace{3.2cm}(e)
\end{center}
\caption{\it
One-loop diagrams that give non-zero contribution to the collinear matrix element $J_n^{\m\n}$, which enters in the definition of the gluon TMDPDF $G_{g/A}^{\m\n [O]}$.
Those diagrams correspond as well to the collinear gluon PDF $f_{g/A}$.
Hermitian conjugates of diagrams (a), (b) and (d) are not shown.
Double lines stand for collinear Wilson lines.
The blob in diagram (a) represents the WFR.
}
\label{fig:collinear}
\end{figure}

In this appendix we present the calculation of the unpolarized gluon TMDPDFs at ${\cal O}(\as)$, using dimensional regularization with the $\overline{\rm MS}$-scheme ($\m^2\to\m^2 e^{\g_E}/(4\pi)$) for ultra-violet (UV) divergences and the $\D$-regulator~\cite{GarciaEchevarria:2011rb} for IR and rapidity divergences.
We use the Keldysh formalism to perform the calculation (see, e.g., \cite{Becher:2007ty,Belitsky:1997ay}).
Our first goal is to show explicitly the cancellation of rapidity divergences in the properly defined gluon TMDPDFs in Eq.~\eqref{eq:tmdsdefinition}.
On the other hand, we will extract the Wilson matching coefficients of the TMDPDF onto its collinear counterparts, as they appear in Eq.~\eqref{eq:opeunpol}.

With the $\D$-regulator, we write the poles of the gluon propagators that involve $p$ or $\bp$ with a real and positive parameters $\D^\pm$,
\begin{align}\label{eq:fermionsDelta}
\frac{-ig^{\m\n}}{(p+k)^2+i0} &\longrightarrow
\frac{-ig^{\m\n}}{(p+k)^2+i\D^-}\,,
\nn\\
\quad\quad
\frac{-ig^{\m\n}}{(\bp+k)^2+i0} &\longrightarrow
\frac{-ig^{\m\n}}{(\bp+k)^2+i\D^+}\,,
\end{align}
and for collinear and soft Wilson lines one has
\begin{align}\label{eq:deltas}
\frac{1}{k^{\pm}\pm i0} \longrightarrow
\frac{1}{k^{\pm}\pm i\d^{\pm}}
\,.
\end{align}
Now, given the fact that the soft and collinear matrix elements must reproduce the soft and collinear limits of full QCD, they need to be regulated consistently, and thus $\d^\pm$ are related with $\D^\pm$ through the large components of the collinear fields,
\begin{align} \label{eq:regul_DeltaDY}
\d^+ = \frac{\D^+}{\bp^-}\,, \quad &\quad \quad \d^- = \frac{\D^-}{p^+} 
\,.
\end{align}
Note that $\D^\pm$ (and hence $\d^\pm$) are regulator parameters, and are set to zero unless they regulate any divergence.

Let us now proceed with the partonic calculation, using Eq.~\eqref{eq:tmdsdefinition2}.
If we consider a hadron with definite helicity $\l$ and take into account only the functions $f_1^f$, $h_1^{\perp g}$ and $g_{1L}^g$ in Eq.~\eqref{eq:decomposition} we then have
\begin{align}
\label{eq:gmunu}
G^{\m\n}(\l) &= 
-\frac{g_\perp^{\m\n}}{2} f_1^g
+ \frac{1}{2}\le(g_\perp^{\m\n} - \frac{2k_{n\perp}^\m k_{n\perp}^\n}{k_{nT}^2}\ri) h_1^{\perp g}
- i\l \frac{\e_{\perp}^{\m\n}}{2} g_{1L}^g 
\,.
\end{align} 
When calculating in $d=4-2\e$ dimensions the tensor structures for the unpolarized and linearly polarized TMDPDFs in \eq{eq:gmunu}, the $d$-dimensional analogues can be obtained by the replacement $\frac{1}{2}(-g^{\m\n}_\perp)\rightarrow \frac{1}{d-2}(-g^{\m\n}_\perp)$. 
For the epsilon tensor multiplying the helicity TMDPDF the situation is more involved and we further discuss this in the calculation in Appendix~\ref{app:helicitynlo}.

At tree level the (naive) collinear matrix element is
\begin{align}
J_{0} &=
\frac{x p^+}{2} \int\frac{dy^-d^2\vecb y_\perp}{(2\pi)^3}\,
e^{-i\le(\frac{1}{2}x y^-p^+-\vecbe y_\perp\cd\vecbe k_{n\perp}\ri)}\,
(-g_\perp^{\m\n}) \frac{1}{d-2}\sum_{pols} \e_{\m}(p) \e_{\n}^*(p)\,
e^{i\frac{1}{2}y^-p^+}
\nn\\
&=
\d(1-x) \d^{(2)}(\vecb k_{n\perp})
\,,
\end{align}

where we have averaged over the gluon polarizations and projected with $g_\perp^{\m\n}$ to extract the unpolarized TMDPDF.

The Wave Function Renormalization (WFR) diagram~\ref{fig:collinear}a and its Hermitian conjugate give
\begin{align}\label{1a}
J_{1}^{(\ref{fig:collinear}a)+(\ref{fig:collinear}a)^*} &=
\d(1-x)\d^{(2)}(\vecb k_{n\perp})
\frac{1}{d-2}\sum_{pols} \e_{\a}(p) \e_{\n}^*(p) 
(-g^{\perp\n}_\m) \frac{-i}{p^2} \frac{1}{2}
\le(n_f\,i\Pi_q^{\a\m} + i\Pi_g^{\a\m} + i\Pi_G^{\a\m} \ri) + h.c.
\nn\\
&=
\frac{\as C_A}{4\pi} \d(1-x)\d^{(2)}(\vecb k_{n\perp})
\bigg[
\le(\frac{1}{\veuv}+\ln\frac{\m^2}{\D^-}\ri) \le(\frac{5}{3}-\frac{2}{3}\frac{n_f}{C_A} \ri)
+ \frac{16}{9} - \frac{4}{9}\frac{n_f}{C_A}\bigg]
\,,
\end{align}
where
\begin{align}
i\Pi_q^{\a\m} &=
-2g^2 T_F \m^{2\ve} \int\frac{d^dk}{(2\pi)^d}
\frac{{\rm Tr}\le[\g^\m\kslash\g^\a(\pslash-\kslash)\ri]}
{[k^2+i0][(p-k)^2+i\D^-]}
\,,\nn\\
i\Pi_g^{\a\m} &=
\frac{1}{2}g^2 C_A \m^{2\ve} \int\frac{d^dk}{(2\pi)^d}
\big[
-g^{\m\g}(p+k)^\d - g^{\g\d}(p-2k)^\m + g^{\d\m}(2p-k)^\g
\big]
\nn\\
&\times
\big[
g^\a_{\;\d}(2p-k)_\g - g_{\d\g}(p-2k)^\a - g^\a_{\;\g}(p+k)_\d
\big]
\frac{1}{[k^2+i0][(p-k)^2+i\D^-]}
\,,\nn\\
\Pi_G^{\a\m} &=
g^2 C_A\m^{2\ve} \int\frac{d^dk}{(2\pi)^d}
\frac{(p-k)^\m k^\a}{[k^2+i0][(p-k)^2+i\D^-]}
\,.
\end{align}

All tadpole diagrams are identically $0$, since $n^2=\nb^2=0$ and they will not be considered any further.
Diagram~\ref{fig:collinear}b and its Hermitian conjugate give
\begin{align}\label{eq:1b}
J_{1}^{(\ref{fig:collinear}b)+(\ref{fig:collinear}b)^*}&=
-ig^2C_A \d(1-x)\d^{(2)}(\vecb k_{n\perp}) \m^{2\e} \int \frac{d^dk}{(2\pi)^d}
\frac{2p^++k^+}{[k^+-i\d^+][(p+k)^2+i\D^-][k^2+i0]}
+ h.c.
\nn\\
&=
\frac{\a_s C_A}{2\pi}
\d(1-x)\d^{(2)}(\vecb k_{n\perp})
\left[
\frac{2}{\veuv}\ln\frac{\d^+}{p^+} + \frac{1}{\veuv} + \ln\frac{\m^2}{\D^-} 
+ 2\ln\frac{\d^+}{p^+}\ln\frac{\m^2}{\D^-} - \ln^2\frac{\d^+}{p^+} + 1 - \frac{7\pi^2}{12}
\right]
\,.
\end{align}

Diagram~(\ref{fig:collinear}c) gives
\begin{align}\label{eq:1c}
J_{1}^{(\ref{fig:collinear}c)}&=
xp^+\pi C_A g^2 \frac{1}{d-2} (-g_{\perp\a\n}) 
\mu^{2\ve}\int \frac{d^dk}{(2\pi)^d}
\frac{\theta(k^+)\d(k^+-(1-x)p^+)\d(k^2)\d^{(2)}(\vecb k_\perp+\vecb k_{n\perp})}
{[(p-k)^2+i\D^-][(p-k)^2-i\D^-]}
(-g_{\perp\l\s})
\nn\\
&\times
\le(-g_{\perp\d\r}+\frac{\bn_\d k_{\perp\r}}{k^++i\d^+}
+\frac{\bn_\r k_{\perp\d}}{k^+-i\d^+} - 
\frac{\bn_\d\bn_\r k_\perp^2}{(k^+)^2+(\d^+)^2}\ri)
\bigg[
g_\perp^{\s\b}-\frac{(p-k)_\perp^\s\bn^\b}{p^+-k^+}
\bigg]
\bigg[
g_\perp^{\l\m} - \frac{(p-k)_\perp^\l\bn^\m}{p^+-k^+}
\bigg]
\nn\\
&\times
\big[
-g^{\n}_{\;\b}(2p-k)^\r + g^{\;\r}_{\b}(p-2k)^\n + g^{\r\n}(p+k)_\b
\big]
\big[
g^{\d}_{\;\m}(p-2k)^\a - g^{\;\a}_\m(2p-k)^\d + g^{\a\d}(p+k)_\m
\big]
\nn\\
&=
\frac{\as C_A}{\pi^2} \bigg[
\frac{x}{1-x}+\frac{(1-x)(1+x^2)}{x}
\bigg]
\frac{(1-x)^2}{[(1-x)^2+(\d^+/p^+)^2]}
\frac{k_{nT}^2}{[k_{nT}^2-i(1-x)\D^-][k_{nT}^2+i(1-x)\D^-]}
\,.
\end{align}

Now we list the Fourier transforms of the previous results:
\begin{align}
\tilde{J}_{0} &=
\d(1-x)
\,,
\end{align}

\begin{align}
{\tilde J}_{1}^{(\ref{fig:collinear}a)+(\ref{fig:collinear}a)^*} &=
\frac{\as C_A}{4\pi} \d(1-x)
\bigg[
\le(\frac{1}{\veuv}+\ln\frac{\m^2}{\D^-}\ri) \le(\frac{5}{3}-\frac{2}{3}\frac{n_f}{C_A} \ri)
+ \frac{16}{9} - \frac{4}{9}\frac{n_f}{C_A}\bigg]
\,,
\end{align}

\begin{align}
\tilde {J}_{1}^{(\ref{fig:collinear}b)+(\ref{fig:collinear}b)^*}&=
\frac{\a_s C_A}{2\pi}
\d(1-x)
\left[
\frac{2}{\veuv}\ln\frac{\d^+}{p^+} + \frac{1}{\veuv} + \ln\frac{\m^2}{\D^-} 
+ 2\ln\frac{\d^+}{p^+}\ln\frac{\m^2}{\D^-} - \ln^2\frac{\d^+}{p^+} + 1 - \frac{7\pi^2}{12}
\right]
\,,
\end{align}

\begin{align}
\tilde{J}_{1}^{(\ref{fig:collinear}c)}&=
\frac{\as C_A}{\pi} \bigg\{
\bigg[
\frac{x}{(1-x)_+}+\frac{(1-x)(1+x^2)}{x} - \d(1-x)\ln\frac{\d^+}{p^+}
\bigg] \le(-L_T+\ln\frac{\m^2}{\D^-}\ri)
\nn\\
&
-\frac{(1-x)(1+x^2)}{x}\ln(1-x) 
- x\le(\frac{\ln(1-x)}{1-x}\ri)_+
+\frac{1}{2}\bigg(\ln^2\frac{\d^+}{p^+}+\frac{\pi^2}{12}\bigg)\d(1-x)
\bigg\}
\,,
\end{align}
where $L_T=\ln(\m^2b_T^2e^{2\g_E}/4)$.

We have used the following identity in $d=2-2\ve$ to perform the Fourier transforms:
\begin{align}
\int d^d\vecb k_\perp e^{i\vecbe k_\perp\cdot \vecbe b_\perp}
f(k_T)
&=
b_T^{-d} (2\pi)^\frac{d}{2} \int_0^\infty dy\, y^\frac{d}{2} J_{\frac{d}{2}-1}(y)\,
f\left(\frac{y}{b_T}\right)\,,
\end{align}
with the particular result
\begin{align}
\int d^d\vecb k_\perp e^{i\vecbe k_\perp\cdot \vecbe b_\perp}
\frac{k_T^2}{k_T^4+\L^4}
&=
\pi\, \ln\frac{4e^{-2\g_E}}{\L^2 b_T^2}
\,,
\end{align}
when $\L\to 0$.
We have also used the following relations:
\begin{align}
f(x)\left[\frac{1}{(1-x)-i\d^+/p^+}
+\frac{1}{(1-x)+i\d^+/p^+}\right] &=
f(x)\le[\frac{2}{(1-x)_+} - 2\ln\frac{\d^+}{p^+}\d(1-x)\ri]
\,,
\nn\\
f(x)\left[\frac{\ln(1-x)}{(1-x)-i\d^+/p^+}
+\frac{\ln(1-x)}{(1-x)+i\d^+/p^+}\right] &=
f(x)\bigg[2\le(\frac{\ln(1-x)}{1-x}\ri)_+ 
-\le(\ln^2\frac{\d^+}{p^+}+\frac{\pi^2}{12}\ri)\d(1-x)
\bigg]
\,,
\end{align}
where $f(x)$ is any function regular at $x\to 1$.

Thus, in IPS, the collinear matrix element for the partonic channel of a gluon splitting into a gluon is
\begin{align}
\label{eq:j1gg}
{\tilde J}_{1}^{g/g} &= 
\frac{\as}{2\pi} \Bigg[  
\d(1-x) \left( \frac{\b_0}{2\veuv}+\frac{2C_A}{\veuv}\ln\frac{\D^+}{Q^2}\right)
\nn\\
& \quad 
+ 2C_A\d(1-x) L_T \ln\frac{\Delta^+}{Q^2}
- L_T\le(P_{g/g}-\d(1-x)\frac{\b_0}{2}\ri) 
\nn\\
& \quad 
+ \ln\frac{\mu^2}{\Delta^-} P_{g/g}
- 2C_A \ln(1-x)\frac{(1-x)(1+x^2)}{x}
- 2C_A x\left(\frac{\ln(1-x)}{1-x}\right)_+
\nn\\
& \quad
+ \d(1-x) \left(-\frac{\pi^2}{2}C_A + \frac{17}{9}C_A - \frac{2}{9}n_f \right) \Bigg]
\,.
\end{align}
The mixed divergences in the result above ($\frac{1}{\veuv}\ln\D^+$) are rapidity divergences, which need to be eliminated by combining it with the soft function as in Eq.~\eqref{eq:tmdsdefinition2} in order to get a well-defined TMDPDF.

\begin{figure}[t]
\begin{center}
\includegraphics[width=0.4\textwidth]{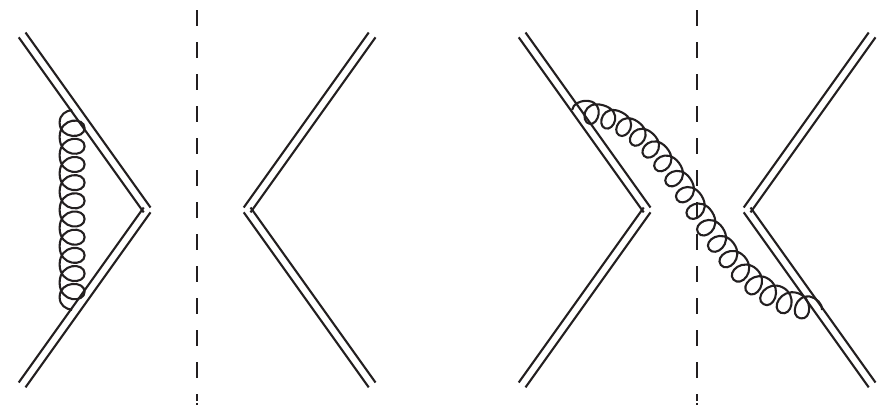}
\\
\vspace{0.2cm}
(a)\hspace{3.5cm}(b)
\end{center}
\caption{\it
One-loop diagrams that give non-zero contribution to the soft function. 
Hermitian conjugate of diagrams (a) and (b) are not shown. 
Double lines stand for soft Wilson lines.}
\label{fig:soft}
\end{figure}

Now we turn our attention to the soft function.
Diagram~(\ref{fig:soft}a) and its Hermitian conjugate give
\begin{align}
S^{(\ref{fig:soft}a)+(\ref{fig:soft}a)^*}&=
-2ig^2 C_A \d^{(2)}(\vecb k_{s\perp}) \mu^{2 \eps}
\int \frac{d^d k}{(2 \pi)^d} \frac{1}{[k^+-i\d^+] [k^-+i\d^-] [k^2+i0]} +h.c.
\nn \\
&=
- \frac{\alpha_s C_A}{2\pi}
\d^{(2)}(\vecb k_{s\perp})
\left[\frac{2}{\veuv^2}-\frac{2}{\veuv}\ln\frac{\d^+\d^-}{\mu^2}+
\ln^2\frac{\d^+\d^-}{\mu^2}+\frac{\pi^2}{2}\right]\, .
\end{align}

Diagram~(\ref{fig:soft}b) and its Hermitian conjugate give
\begin{align}
S^{(\ref{fig:soft}b)+(\ref{fig:soft}b)^*}&=
-4\pi g^2 C_A  \mu^{2\ve}\int\frac{d^dk}{(2\pi)^d}
\frac{\d^{(2)}(\vecb k_\perp+\vecb k_{s\perp}) \d(k^2)\theta(k^+)}{[k^++i\d^+][-k^-+i\d^-]} + h.c.
\nn\\
&=
-\frac{\a_s C_A}{\pi^2}
\frac{1}{k_{sT}^2-\d^+\d^-} \ln\frac{\d^+\d^-}{k_{sT}^2}
\,.
\end{align}

Using the relation
\begin{align}
\int d^d\vecb k_\perp e^{i\vecbe k_\perp\cdot \vecbe b_\perp}
f(k_T)
&=
b_T^{-d} (2\pi)^\frac{d}{2} \int_0^\infty dy\, 
y^\frac{d}{2} J_{\frac{d}{2}-1}(y)\,
f(y/b_T)
\end{align}
and the particular result
\begin{align}
\int d^d\vecb k_\perp e^{i\vecbe k_\perp\cdot \vecbe b_\perp}
\frac{1}{k_{T}^2-\L^2} \ln\frac{\L^2}{k_{T}^2}
&=
\pi \left(
-\frac{1}{2}\ln^2\frac{4e^{-2\g_E}}{\L^2 b_T^2} - \frac{\pi^2}{3}
\right)
\end{align}
when $\L\to 0$, we combine the virtual and real contributions to the soft function in IPS to get
\begin{align}\label{eq:softfirstorder}
\tilde{S}_1\le(\frac{\D^-}{p^+},\frac{\D^+}{\bp^-}\ri) &=
\frac{\as C_A}{2\pi} \le[
- \frac{2}{\veuv^2} + \frac{2}{\veuv} \ln\frac{\D^-\D^+}{\m^2Q^2}
+ L_T^2 + 2L_T\ln\frac{\D^-\D^+}{\m^2Q^2} + \frac{\pi^2}{6}
\ri]
\,,
\end{align}
where we have made the replacements $\d^+=\D^+/\bp^-$ and $\d^-=\D^-/p^+$.

Combining the collinear and soft matrix elements as in Eq.~\eqref{eq:tmdsdefinition2} we get the unpolarized TMDPDF in IPS for the $g/g$ channel:
\begin{align}\label{eq:unpoltmdnlo}
f^g_{1 g/g}(x,b_T;\mu,\zeta) & =
\d(1-x) + \le[
{\tilde J}^{g/g}_{1} 
- \frac{1}{2}\d(1-x) {\tilde S}_1\le(\frac{1}{\a}\frac{\D^+}{p^+},\frac{\D^+}{\bp^-}\ri)
\ri]
\nn\\
&=
\d(1-x) + \frac{\alpha_s}{2\pi} 
	\Bigg\{ \left[ \frac{C_A}{\veuv^2} + \frac{1}{\veuv} \left( \frac{\beta_0}{2}+C_A \ln \frac{\mu^2}{\z} \right) \right] \delta(1-x) \Bigg.
\nonumber\\
&\quad 
+ C_A\d(1-x)\le( - \frac{1}{2}L_T^2 + L_T\ln\frac{\m^2}{\z} - \frac{\pi^2}{12}\ri)\delta(1-x) 
- L_T\le(P_{g/g}-\d(1-x)\frac{\b_0}{2}\ri) 
\nonumber\\
&\quad 
+ \ln \frac{\mu^2}{\Delta^-} P_{g/g} - 2C_A \ln(1-x) \frac{(1-x)(1+x^2)}{x} 
- 2C_A x\left( \frac{\ln(1-x)}{1-x}\right)_+ 
\nonumber\\
&\quad 
\Bigg.
+ \d(1-x) \left(-\frac{\pi^2}{2}C_A + \frac{17}{9}C_A - \frac{2}{9}n_f \right) 
\Bigg\}
\,.
\end{align}
Notice that in this result the rapidity divergences have disappeared, and that we have ended up with UV poles, which will give us the QCD evolution of this quantity, and a single IR pole (parametrized by the $\D^-$), which is a manifestation of true long-distance physics of QCD or confinement.

Finally we calculate the $g/q$ channel, which corresponds to diagram~\ref{fig:collinear}e:
\begin{align}\label{eq:1e}
J_{1}^{(\ref{fig:collinear}e)}&=
xp^+\pi C_F g^2 \mu^{2\ve} \int \frac{d^dk}{(2\pi)^d}
\frac{\theta(k^+)\d(k^+-(1-x)p^+)\d(k^2)\d^{(2)}(\vecb k_\perp+\vecb k_{n\perp})}
{[(p-k)^2+i\D^-][(p-k)^2-i\D^-]}
\nn\\
&\times
{\rm Tr}\big[
\pslash\g^\b\kslash\g^\a
\big]
\le(g_\perp^{\l\a}-\frac{(p-k)_\perp^\l\bn^\a}{p^+-k^+}\ri)
\le(g_\perp^{\r\b}-\frac{(p-k)_\perp^\r\bn^\b}{p^+-k^+}\ri)
\le(-g_{\perp\l\r}\ri)
\nn\\
&=
\frac{\as}{2\pi^2} {\cal P}_{g/q}
\frac{k_{nT}^2}{k_{nT}^4+(1-x)^2(\D^-)^2}
\,.
\end{align}

In IPS we get
\begin{align}
{\tilde f}^g_{1 g/q}(x,b_T;\mu,\zeta)&=
\frac{\as}{2\pi} {\cal P}_{g/q}
\le(-L_T+\ln\frac{\m^2}{\D^-} - \ln(1-x) \ri)
\,.
\end{align}
This channel at this order does not have any rapidity divergences.

Next we calculate the collinear unpolarized gluon PDF, since our goal is to obtain the OPE Wilson coefficient of the perturbative expansion of the unpolarized gluon TMDPDF at large transverse momentum.
Virtual diagrams are the same as for the unpolarized gluon TMDPDF, apart from the $\d^{(2)}(\vecb k_{n\perp})$.
The Wave Function Renormalization (WFR) diagram~\ref{fig:collinear}a and its Hermitean conjugate give
\begin{align}
f_{1}^{g(\ref{fig:collinear}a)+(\ref{fig:collinear}a)^*} &=
\frac{\as C_A}{4\pi} \d(1-x)
\bigg[
\le(\frac{1}{\veuv}+\ln\frac{\m^2}{\D^-}\ri) \le(\frac{5}{3}-\frac{2}{3}\frac{n_f}{C_A} \ri)
+ \frac{16}{9} - \frac{4}{9}\frac{n_f}{C_A}\bigg]
\,.
\end{align}

Diagram~\ref{fig:collinear}b and its Hermitian conjugate give
\begin{align}
f_{1}^{g(\ref{fig:collinear}b)+(\ref{fig:collinear}b)^*}&=
\frac{\a_s C_A}{2\pi}\d(1-x)
\left[
\frac{2}{\veuv}\ln\frac{\d^+}{p^+} + \frac{1}{\veuv} + \ln\frac{\m^2}{\D^-} 
+ 2\ln\frac{\d^+}{p^+}\ln\frac{\m^2}{\D^-} - \ln^2\frac{\d^+}{p^+} + 1 - \frac{7\pi^2}{12}
\right]
\,.
\end{align}

Diagram~(\ref{fig:collinear}c) gives
\begin{align}
\label{eq:f1g7c}
f_{1}^{g(\ref{fig:collinear}c)}&=
xp^+\pi C_A g^2 \frac{1}{d-2} (-g_{\perp\a\n}) \m^{2\e} \int \frac{d^dk}{(2\pi)^d}
\frac{\theta(k^+)\d(k^+-(1-x)p^+)\d(k^2)}
{[(p-k)^2+i\D^-][(p-k)^2-i\D^-]}
(-g_{\perp\l\s})
\nn\\
&\times
\le(-g_{\perp\d\r}+\frac{\bn_\d k_{\perp\r}}{k^++i\d^+}
+\frac{\bn_\r k_{\perp\d}}{k^+-i\d^+} - 
\frac{\bn_\d\bn_\r k_\perp^2}{(k^+)^2+(\d^+)^2}\ri)
\bigg[
g_{\perp\b}^{\s}-\frac{(p-k)_\perp^\s\bn_\b}{p^+-k^+}
\bigg]
\bigg[
g_{\perp\m}^{\l} - \frac{(p-k)_\perp^\l\bn_\m}{p^+-k^+}
\bigg]
\nn\\
&\times
\big[
-g^{\n\b}(2p-k)^\r + g^{\b\r}(p-2k)^\n + g^{\r\n}(p+k)^\b
\big]
\big[
g^{\d\m}(p-2k)^\a - g^{\m\a}(2p-k)^\d + g^{\a\d}(p+k)^\m
\big]
\nn\\
&=
\frac{\as C_A}{\pi} \bigg\{
\bigg[
\frac{x}{(1-x)_+}+\frac{(1-x)(1+x^2)}{x} - \d(1-x)\ln\frac{\d^+}{p^+}
\bigg] \le(\frac{1}{\veuv}+\ln\frac{\m^2}{\D^-}\ri)
\nn\\
&
-\frac{(1-x)(1+x^2)}{x}\ln(1-x) 
- x\le(\frac{\ln(1-x)}{1-x}\ri)_+
+\frac{1}{2}\bigg(\ln^2\frac{\d^+}{p^+}+\frac{\pi^2}{12}\bigg)\d(1-x)
\bigg\}
\,.
\end{align}

The unpolarized collinear gluon PDF in the $g/g$ channel is then given by
\begin{align}
\label{eq:f1ggg}
f^g_{1 g/g}(x;\mu) &= 
\d(1-x) + \frac{\as}{2\pi} \Bigg[ \left( \frac{1}{\veuv} + \ln\frac{\mu^2}{\Delta^-} \right) P_{g/g} - 2C_A \ln(1-x)\frac{(1-x)(1+x^2)}{x} 
- 2C_A x\left( \frac{\ln(1-x)}{1-x}\right)_+ \Bigg.
\nonumber\\
&\quad 
\Bigg.
+ \d(1-x) \left(-\frac{\pi^2}{2}C_A + \frac{17}{9}C_A - \frac{2}{9}n_f \right)\Bigg]
\end{align}
Notice that the single IR pole, which is parametrized by the logarithm of $\D^-$, is the true collinear divergence of the PDF, remnant of QCD long-distance physics. 

We are ready now, given Eq.~\eqref{eq:opeunpol}, to extract the matching coefficient of the TMDPDF onto the PDF in the $g/g$ channel:
\begin{align}
\tilde{C}^f_{g/g} &= 
\d(1-x) + \frac{\as}{2\pi} \left[ 
C_A\d(1-x)\le( - \frac{1}{2}L_T^2 + L_T\ln\frac{\m^2}{Q^2} - \frac{\pi^2}{12}\ri)
- L_T\le(P_{g/g}-\d(1-x)\frac{\b_0}{2}\ri) 
\right]
\,.
\end{align}

For the $g/q$ channel the unpolarized collinear gluon PDF is given by
\begin{align}
f^{g (\ref{fig:collinear}e)}_{1 g/q}(x;\mu)  &=
xp^+\pi C_F g^2 \int \mu^{2\ve} \frac{d^dk}{(2\pi)^d}
\frac{\theta(k^+)\d(k^+-(1-x)p^+)\d(k^2)}
{[(p-k)^2+i\D^-][(p-k)^2-i\D^-]}
\nn\\
&\times
{\rm Tr}\big[
\pslash\g_\b\kslash\g_\a
\big]
\le(g_\perp^{\l\a}-\frac{(p-k)_\perp^\l\bn^\a}{p^+-k^+}\ri)
\le(g_\perp^{\r\b}-\frac{(p-k)_\perp^\r\bn^\b}{p^+-k^+}\ri)
\le(-g_{\perp\l\r}\ri)
\nn\\
&=
\frac{\as}{2\pi} \bigg[
\le(\frac{1}{\veuv}+\ln\frac{\m^2}{\D^-}\ri)
{\cal P}_{g/q}
-{\cal P}_{g/q}\ln(1-x) - C_F x
\bigg]
\,,
\end{align}
and thus the matching of the TMDPDF onto the PDF in the $g/q$ channel is
\begin{align}
\tilde{C}^f_{g/q} &= 
\frac{\as}{2\pi} \bigg[
-L_T {\cal P}_{g/q} + C_F x
\bigg]
\,.
\end{align}

\section{OPE of $h_1^{\perp g}$ at NLO}
\label{app:linearnlo}

The calculation in this Appendix follows the same logic as in the previous one, so we limit ourselves to provide the relevant results.

For the distribution of linearly polarized gluons inside an unpolarized hadron at NLO only real diagrams contribute.
In the $g/g$ channel we have:
\begin{align}
J_{1}^{(\ref{fig:collinear}c)}&=
xp^+\pi C_A g^2 \frac{1}{d-2} (-g_{\perp\a\n}) 
\mu^{2\ve} \int \frac{d^dk}{(2\pi)^d}
\frac{\theta(k^+)\d(k^+-(1-x)p^+)\d(k^2)\d^{(2)}(\vecb k_\perp+\vecb k_{n\perp})}
{[(p-k)^2+i\D^-][(p-k)^2-i\D^-]}
\nn\\
&\times
\frac{d-2}{d-3}\le(\frac{g_{\perp\l\s}}{d-2}-\frac{(p-k)_{\perp\l}(p-k)_{\perp\s}}{(p-k)_\perp^2}\ri)
\nn\\
&\times
\le(-g_{\perp\d\r}+\frac{\bn_\d k_{\perp\r}}{k^++i\d^+}
+\frac{\bn_\r k_{\perp\d}}{k^+-i\d^+} - 
\frac{\bn_\d\bn_\r k_\perp^2}{(k^+)^2+(\d^+)^2}\ri)
\bigg[
g_{\perp\b}^{\s}-\frac{(p-k)_\perp^\s\bn_\b}{p^+-k^+}
\bigg]
\bigg[
g_{\perp\m}^{\l} - \frac{(p-k)_\perp^\l\bn_\m}{p^+-k^+}
\bigg]
\nn\\
&\times
\big[
-g^{\n\b}(2p-k)^\r + g^{\b\r}(p-2k)^\n + g^{\r\n}(p+k)^\b
\big]
\big[
g^{\d\m}(p-2k)^\a - g^{\m\a}(2p-k)^\d + g^{\a\d}(p+k)^\m
\big]
\nn\\
&=
\frac{\as C_A}{\pi^2} \frac{1-x}{x}
\frac{k_{nT}^2}{[k_{nT}^2-i(1-x)\D^-][k_{nT}^2+i(1-x)\D^-]}
\,.
\end{align}
Notice that in the projector in the second line, we have replaced a factor $2$ by $d-2$, which makes it orthogonal to the one for the unpolarized TMDPDF ($g_{\m\n}^\perp$ in \eq{eq:gmunu}) in $d$ dimensions. 
While the $(d-2)/(d-3)$ factor is a normalization factor necessary to single out the linearly polarized TMDPDF.
However, this is not relevant at this perturbative order, since there are no poles in $\e$.

For the $g/q$ channel we have:
\begin{align}
J_{1}^{(\ref{fig:collinear}e)}&=
xp^+\pi C_F g^2 \mu^{2\ve} \int \frac{d^dk}{(2\pi)^d}
\frac{\theta(k^+)\d(k^+-(1-x)p^+)\d(k^2)\d^{(2)}(\vecb k_\perp+\vecb k_{n\perp})}
{[(p-k)^2+i\D^-][(p-k)^2-i\D^-]}
\nn\\
&\times
{\rm Tr}\big[
\pslash\g_\b\kslash\g_\a
\big]
\le(g_\perp^{\l\a}-\frac{(p-k)_\perp^\l\bn^\a}{p^+-k^+}\ri)
\le(g_\perp^{\r\b}-\frac{(p-k)_\perp^\r\bn^\b}{p^+-k^+}\ri)
\frac{d-2}{d-3}\le(\frac{g_{\perp\l\r}}{d-2}-\frac{(p-k)_{\perp\l}(p-k)_{\perp\r}}{(p-k)_\perp^2}\ri)
\nn\\
&=
\frac{\as C_F}{\pi^2} \frac{1-x}{x}
\frac{k_{nT}^2}{[k_{nT}^2-i(1-x)\D^-][k_{nT}^2+i(1-x)\D^-]}
\,.
\end{align}

In order to go to IPS we do
\begin{align}
{\tilde h}_1^{\perp g\,(2)}(x,b_T) &=
-2\pi\int dk_{nT}\, k_{nT}\, J_2(k_{nT}b_T)\, h_1^{\perp g}(x,k_{nT})
\,,
\end{align}
so then
\begin{align}
{\tilde h}_{1 g/g}^{\perp g\,(2)} &=
-\frac{\as C_A}{\pi} \frac{1-x}{x}
\,,
\end{align}
and
\begin{align}
{\tilde h}_{1 g/q}^{\perp g\,(2)} &=
-\frac{\as C_F}{\pi} \frac{1-x}{x}
\,.
\end{align}
Notice that at this perturbative order we do not find rapidity divergences.

The matchings of the linearly polarized gluon TMDPDF onto the collinear PDFs in the $g/g$ and $g/q$ channels are then:
\begin{align}
\tilde{C}^h_{g/g} &= 
-\frac{\as}{\pi} C_A \frac{1-x}{x}
\,,\nn\\
\tilde{C}^h_{g/q} &= 
-\frac{\as}{\pi} C_F \frac{1-x}{x}
\,.
\end{align}
Those results follow directly from the OPE of ${\tilde h}_{1}^{\perp g\,(2)}$ in terms of the collinear quark/gluon PDFs, given that at LO the later are simply $\d(1-x)$ while ${\tilde h}_{1}^{\perp g\,(2)}$ starts at order $\as$.

\section{OPE of $g_{1L}^g$ at NLO}
\label{app:helicitynlo}

Again the calculation in this Appendix follows the same logic as in the previous ones, so we limit ourselves to provide the relevant results. 
For the gluon in a gluon helicity matching coefficient there are always two epsilon tensors in each diagram calculation. 
This product can be rewritten in terms of metric tensors and we use a product of two transverse epsilon tensors in $d$-dimensions with the normalization 
\begin{align}
\e^\perp_{\m\n} \e_\perp^{\m\n} &= (d-2)(d-3)
\,.
\end{align}
For the calculation of the gluon in a quark we have to calculate a trace containing one $\gamma_5$. 
This is done through the identification 
$\text{Tr}(\g^\m \g^\n \g^\a \g^b\g_5) \rightarrow - 4i\e^{\m\n\a\b}$ 
in order to once again obtain the product of two epsilon tensors. 
For a calculation of the matching coefficients in a different scheme and a more careful treatment of $\gamma_5$ and the epsilon tensor, we refer the reader to \cite{BufDieKas}.

At tree level the (naive) collinear matrix element is
\begin{align}
J_{0} &=
\frac{x p^+}{2} \int\frac{dy^-d^2\vecb y_\perp}{(2\pi)^3}\,
e^{-i\le(\frac{1}{2}x y^-p^+-\vecbe y_\perp\cd\vecbe k_{n\perp}\ri)}\,
\frac{(i\e_\perp^{\m\n})(-i\e_{\m\n}^\perp)}{(d-2)(d-3)}
e^{i\frac{1}{2}y^-p^+}
\nn\\
&=
\d(1-x) \d^{(2)}(\vecb k_{n\perp})
\,.
\end{align}

The Wave Function Renormalization (WFR) diagram~\ref{fig:collinear}a and its Hermitean conjugate give
\begin{align}
J_{1}^{(\ref{fig:collinear}a)+(\ref{fig:collinear}a)^*} &=
\d(1-x)\d^{(2)}(\vecb k_{n\perp})
\frac{(-i\e^\perp_{\a\n})}{(d-2)(d-3)}
(i\e^{\perp\n}_{\m})
 \frac{-i}{p^2} \frac{1}{2}
\le(n_f\,i\Pi_q^{\a\m} + i\Pi_g^{\a\m} + i\Pi_G^{\a\m} \ri) + h.c.
\nn\\
&=
\frac{\as C_A}{4\pi} \d(1-x)\d^{(2)}(\vecb k_{n\perp})
\bigg[
\le(\frac{1}{\veuv}+\ln\frac{\m^2}{\D^-}\ri) \le(\frac{5}{3}-\frac{2}{3}\frac{n_f}{C_A} \ri)
+ \frac{16}{9} - \frac{4}{9}\frac{n_f}{C_A}\bigg]
\,.
\end{align}

Diagram~\ref{fig:collinear}b and its Hermitean conjugate give
\begin{align}
J_{1}^{(\ref{fig:collinear}b)+(\ref{fig:collinear}b)^*}&=
-ig^2C_A \d(1-x)\d^{(2)}(\vecb k_{n\perp}) \m^{2\e} \int \frac{d^dk}{(2\pi)^d}
\frac{2p^++k^+}{[k^+-i\d^+][(p+k)^2+i\D^-][k^2+i0]}
+ h.c.
\nn\\
&=
\frac{\a_s C_A}{2\pi}
\d(1-x)\d^{(2)}(\vecb k_{n\perp})
\left[
\frac{2}{\veuv}\ln\frac{\d^+}{p^+} + \frac{1}{\veuv} + \ln\frac{\m^2}{\D^-} 
+ 2\ln\frac{\d^+}{p^+}\ln\frac{\m^2}{\D^-} - \ln^2\frac{\d^+}{p^+} + 1 - \frac{7\pi^2}{12}
\right]
\,.
\end{align}

Diagram~(\ref{fig:collinear}c) gives
\begin{align}
J_{1}^{(\ref{fig:collinear}c)}&=
xp^+\pi C_A g^2  
\frac{(-i\e^\perp_{\a\n})}{(d-2)(d-3)}
\mu^{2\ve}\int \frac{d^dk}{(2\pi)^d}
\frac{\theta(k^+)\d(k^+-(1-x)p^+)\d(k^2)\d^{(2)}(\vecb k_\perp+\vecb k_{n\perp})}
{[(p-k)^2+i\D^-][(p-k)^2-i\D^-]}
(i\e^{\perp}_{\l\s})
\nn\\
&\quad\times
\le(-g_{\perp\d\r}+\frac{\bn_\d k_{\perp\r}}{k^++i\d^+}
+\frac{\bn_\r k_{\perp\d}}{k^+-i\d^+} - 
\frac{\bn_\d\bn_\r k_\perp^2}{(k^+)^2+(\d^+)^2}\ri)
\bigg[
g_{\perp\b}^{\s}-\frac{(p-k)_\perp^\s\bn_\b}{p^+-k^+}
\bigg]
\bigg[
g_{\perp\m}^{\l} - \frac{(p-k)_\perp^\l\bn_\m}{p^+-k^+}
\bigg]
\nn\\
&\quad\times
\big[
-g^{\n\b}(2p-k)^\r + g^{\b\r}(p-2k)^\n + g^{\r\n}(p+k)^\b
\big]
\big[
g^{\d\m}(p-2k)^\a - g^{\m\a}(2p-k)^\d + g^{\a\d}(p+k)^\m
\big]
\nn\\
&=
\frac{\as C_A}{\pi^2} \bigg[
\frac{x}{1-x}+\frac{(1-x)(1+x^2)}{x}
-\frac{(1-x)^3}{x}
\bigg]
\frac{(1-x)^2}{[(1-x)^2+(\d^+/p^+)^2]}
\nn\\
&\quad\times
\frac{k_{nT}^2}{[k_{nT}^2-i(1-x)\D^-][k_{nT}^2+i(1-x)\D^-]}
\,.
\end{align}

Diagram~\ref{fig:collinear}e is given by the $g/q$ channel:
\begin{align}
J_{1}^{(\ref{fig:collinear}e)}&=
xp^+ 2\pi C_F g^2 \mu^{2\ve} \int \frac{d^dk}{(2\pi)^d}
\frac{\theta(k^+)\d(k^+-(1-x)p^+)\d(k^2)\d^{(2)}(\vecb k_\perp+\vecb k_{n\perp})}
{[(p-k)^2+i\D^-][(p-k)^2-i\D^-]}
\nn\\
&\times
\le(i\e^\perp_{\l\r}\ri) \frac{1}{(d-2)(d-3)}
{\rm Tr}\big[
-\pslash\g_5\g_\b\kslash\g_\a
\big]
\le(g_\perp^{\l\a}-\frac{(p-k)_\perp^\l\bn^\a}{p^+-k^+}\ri)
\le(g_\perp^{\r\b}-\frac{(p-k)_\perp^\r\bn^\b}{p^+-k^+}\ri)
\nn\\
&=
\frac{\as}{2\pi^2} {\cal P}_{\D g/\D q}(x)
\frac{k_{nT}^2}{k_{nT}^4+(1-x)^2(\D^-)^2}
\,,
\end{align}
where
\begin{align}
{\cal P}_{\D g/\D q}(x) &=
C_F\frac{1-(1-x)^2}{x}
\,.
\end{align}

Now we list the Fourier transform of the previous results:
\begin{align}
\tilde{J}_{1}^{(\ref{fig:collinear}a)+(\ref{fig:collinear}a)^*} &=
\frac{\as C_A}{4\pi} \d(1-x)
\bigg[
\le(\frac{1}{\veuv}+\ln\frac{\m^2}{\D^-}\ri) \le(\frac{5}{3}-\frac{2}{3}\frac{n_f}{C_A} \ri)
+ \frac{16}{9} - \frac{4}{9}\frac{n_f}{C_A}\bigg]
\,.
\end{align}

\begin{align}
{\tilde J}_{1}^{(\ref{fig:collinear}b)+(\ref{fig:collinear}b)^*}&=
\frac{\a_s C_A}{2\pi}
\d(1-x)
\left[
\frac{2}{\veuv}\ln\frac{\d^+}{p^+} + \frac{1}{\veuv} + \ln\frac{\m^2}{\D^-} 
+ 2\ln\frac{\d^+}{p^+}\ln\frac{\m^2}{\D^-} - \ln^2\frac{\d^+}{p^+} + 1 - \frac{7\pi^2}{12}
\right]
\,.
\end{align}

\begin{align}
{\tilde J}_{1}^{(\ref{fig:collinear}c)}&=
\frac{\as C_A}{\pi} \bigg\{
\bigg[
\frac{x}{(1-x)_+}+\frac{(1-x)(1+x^2)}{x}
-\frac{(1-x)^3}{x} - \d(1-x)\ln\frac{\d^+}{p^+}
\bigg]
\le(-L_T+\ln\frac{\m^2}{\D^-} \ri)
\nn\\
&-
\bigg[ \frac{(1-x)(1+x^2)}{x} - \frac{(1-x)^3}{x} \bigg]
\ln(1-x)
- x\le(\frac{\ln(1-x)}{1-x}\ri)_+
+ \frac{1}{2}\le(\ln^2\frac{\d^+}{p^+}+\frac{\pi^2}{12}\ri)\d(1-x)
\bigg\}
\,.
\end{align}

\begin{align}
{\tilde J}_{1}^{(\ref{fig:collinear}e)}&=
\frac{\as}{2\pi} {\cal P}_{\D g/\D q}(x)
\bigg(
\ln\frac{\m^2}{\D^-} - L_T - \ln(1-x)
\bigg)
\,.
\end{align}

Combining all the results, the collinear matrix element for the partonic channel of a gluon splitting into a gluon is
\begin{align}
{\tilde J}_{1}^{g/g} &= 
\frac{\as}{2\pi} \Bigg[  
\d(1-x) \left( \frac{\b_0}{2\veuv}+\frac{2C_A}{\veuv}\ln\frac{\D^+}{Q^2}\right)
\nn\\
& \quad 
+ 2C_A\d(1-x) L_T \ln\frac{\Delta^+}{Q^2}
- L_T\le(P_{\D g/\D g}-\d(1-x)\frac{\b_0}{2}\ri) 
\nn\\
& \quad 
+ \ln\frac{\mu^2}{\Delta^-} P_{\D g/\D g}
-2C_A\bigg[ \frac{(1-x)(1+x^2)}{x} - \frac{(1-x)^3}{x} \bigg]\ln(1-x)
- 2C_A x \left(\frac{\ln(1-x)}{1-x}\right)_+
\nn\\
& \quad
+ \d(1-x) \left(-\frac{\pi^2}{2}C_A + \frac{17}{9}C_A - \frac{2}{9}n_f \right) \Bigg]
\,.
\end{align}
The mixed divergences in the result above ($\frac{1}{\veuv}\ln\D^+$) are rapidity divergences, which need to be eliminated by combining it with the soft function (from \eq{eq:softfirstorder}) as in Eq.~\eqref{eq:tmdsdefinition2} in order to get a well-defined TMDPDF.
The result is
\begin{align}
{\tilde g}_{1L\,g/g}^g &=
\d(1-x) + \le[
{\tilde J}^{g/g}_{1} 
- \frac{1}{2}\d(1-x) {\tilde S}_1\le(\frac{1}{\a}\frac{\D^+}{p^+},\frac{\D^+}{\bp^-}\ri)
\ri]
\nn\\
&=
\d(1-x) + \frac{\as}{2\pi} 
\Bigg\{
\d(1-x)
\bigg[
\frac{C_A}{\veuv^2}
+\frac{1}{\veuv}\le(
\frac{\b_0}{2}
+
C_A\ln\frac{\m^2}{\z}
\ri)
\bigg]
-L_T\bigg[
{\cal P}_{\D g/\D g}
-\frac{\b_0}{2}\d(1-x)
\bigg]
\nn\\
&
+ C_A\d(1-x)\le( - \frac{1}{2}L_T^2 + L_T\ln\frac{\m^2}{\z} - \frac{\pi^2}{12}\ri)\d(1-x)
+{\cal P}_{\D g/\D g} \ln\frac{\m^2}{\D}
+\d(1-x)\bigg[\frac{17}{9}C_A - \frac{2}{9}n_f\bigg]
\nn\\
&
-2C_A\bigg[ \frac{(1-x)(1+x^2)}{x} - \frac{(1-x)^3}{x} \bigg]\ln(1-x)
- 2C_A x \le(\frac{\ln(1-x)}{1-x}\ri)_+ 
- C_A\frac{\pi^2}{2}\d(1-x)
\Bigg\}
\,.
\end{align}
Notice that the UV poles coincide with the unpolarized gluon TMDPDF in \eq{eq:unpoltmdnlo}, i.e., as expected, both have the same anomalous dimension.

Next we calculate the collinear gluon helicity.
Virtual diagrams are the same as for the gluon helicity TMDPDF.
The Wave Function Renormalization (WFR) diagram~\ref{fig:collinear}a and its Hermitean conjugate give
\begin{align}
g_{1L}^{g(\ref{fig:collinear}a)+(\ref{fig:collinear}a)^*} &=
\frac{\as C_A}{4\pi} \d(1-x)
\bigg[
\le(\frac{1}{\veuv}+\ln\frac{\m^2}{\D^-}\ri) \le(\frac{5}{3}-\frac{2}{3}\frac{n_f}{C_A} \ri)
+ \frac{16}{9} - \frac{4}{9}\frac{n_f}{C_A}\bigg]
\,.
\end{align}

\begin{align}
g_{1L}^{g(\ref{fig:collinear}b)+(\ref{fig:collinear}b)^*}&=
\frac{\a_s C_A}{2\pi} \d(1-x)
\left[
\frac{2}{\veuv}\ln\frac{\d^+}{p^+} + \frac{1}{\veuv} + \ln\frac{\m^2}{\D^-} 
+ 2\ln\frac{\d^+}{p^+}\ln\frac{\m^2}{\D^-} - \ln^2\frac{\d^+}{p^+} + 1 - \frac{7\pi^2}{12}
\right]
\,.
\end{align}

\begin{align}
g_{1L}^{g(\ref{fig:collinear}c)}&=
\frac{\as C_A}{\pi} \bigg\{
\bigg[
\frac{x}{(1-x)_+}+\frac{(1-x)(1+x^2)}{x} 
-\frac{(1-x)^3}{x}
- \d(1-x)\ln\frac{\d^+}{p^+}
\bigg] \le(\frac{1}{\veuv}+\ln\frac{\m^2}{\D^-}\ri)
\nn\\
&
-\bigg[\frac{(1-x)(1+x^2)}{x} 
-\frac{(1-x)^3}{x}\bigg]\ln(1-x)
- x\le(\frac{\ln(1-x)}{1-x}\ri)_+
+\frac{1}{2}\bigg(\ln^2\frac{\d^+}{p^+}+\frac{\pi^2}{12}\bigg)\d(1-x)
+2(1-x)
\bigg\}
\,.
\end{align}

The collinear gluon helicity PDF in the $g/g$ channel is then
\begin{align}
g_{1L}^{g/g} &=
\d(1-x) + \frac{\as}{2\pi} 
\Bigg\{
{\cal P}_{\D g/\D g}\le(\frac{1}{\veuv}+\ln\frac{\m^2}{\D^-}\ri)
+C_A\d(1-x)\bigg[
\frac{17}{9} - \frac{2}{9}\frac{n_f}{C_A}
\bigg]
\nn\\
&
-2C_A\bigg[\frac{(1-x)(1+x^2)}{x} 
-\frac{(1-x)^3}{x}\bigg]\ln(1-x)
- 2C_A x \le(\frac{\ln(1-x)}{1-x}\ri)_+
-\frac{\pi^2}{2}\d(1-x)C_A
+4C_A(1-x)
\Bigg\}
\,.
\end{align}

The collinear gluon helicity PDF in the $g/q$ channel is given by diagram~\ref{fig:collinear}e:
\begin{align}
g_{1L}^{g(\ref{fig:collinear}e)}&=
\frac{\as}{2\pi} \bigg[
\le(\frac{1}{\veuv}+\ln\frac{\m^2}{\D^-}\ri)
{\cal P}_{\D g/\D q}
-{\cal P}_{\D g/\D q}\ln(1-x) 
+ C_F 2(1-x)
\bigg]
\,.
\end{align}

Thus the matching of the gluon helicity TMDPDF onto the collinear gluon helicity PDF in the ($g/g$) channel is
\begin{align}\label{eq:C_hel_gg}
\tilde{C}^g_{g/g} &= 
\d(1-x) + \frac{\as}{2\pi} \left[ 
C_A\d(1-x)\le( - \frac{1}{2}L_T^2 + L_T\ln\frac{\m^2}{\z} - \frac{\pi^2}{12}\ri)
- L_T\le({\cal P}_{\D g/\D g}-\d(1-x)\frac{\b_0}{2}\ri)
-4C_A(1-x)
\right]
\,.
\end{align}
The matching in the $g/q$ channel is 
\begin{align}\label{eq:C_hel_gq}
\tilde{C}^g_{g/q} &= 
\frac{\as}{2\pi} \bigg[
-L_T {\cal P}_{\D g/\D q} 
- C_F 2(1-x)
\bigg]
\,.
\end{align}

\section{Hard Part at NLO}
\label{app:hard}

\begin{figure}[t]
\begin{center}
\includegraphics[width=0.8\textwidth]{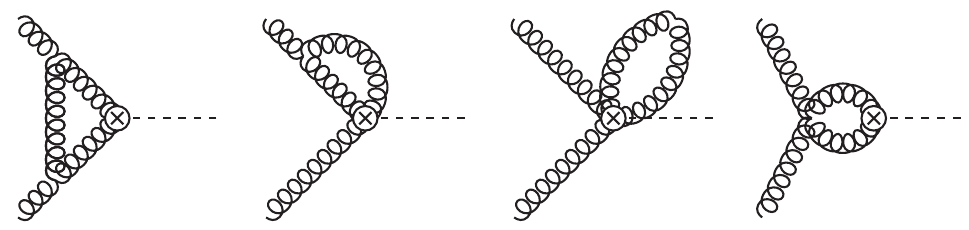}
\\
\vspace{0.2cm}
(a)\hspace{3.2cm}(b)\hspace{3.2cm}(c)\hspace{3.2cm}(d)
\end{center}
\caption{\it
One-loop virtual diagrams for the Higgs production in full QCD.
Hermitian conjugate diagrams are not shown.
The crossed diagram of (a) and the analogous one of (b) are not shown.
}
\label{fig:hard}
\end{figure}

In this appendix we report the explicit NLO calculation of the hard matching coefficient $C_H$ in \eq{eq:hardcoeff}, which accounts for the hard reaction in the gluon-gluon fusion to Higgs boson process.
For simplicity we will take $\D^\pm=\D$.
The tensor structure that appears in the effective $ggH$ vertex $g(p)+g(k)\to H$ is:
\begin{align}
g_{\mu\nu}^H(p,k) = (p\cd k) g_{\mu\nu} - p_\nu k_\mu
\,.
\end{align}

The results of the diagrams in figure~\ref{fig:hard} are, omitting a common prefactor $\d(1-x_A)\d(1-x_B)\d^{(2)}(\vecb q_\perp)$ :
\begin{align}
V^{(\ref{fig:hard}a)} & = 
\frac{-i 2\pi\alpha_s C_A}{(d-2)(p\bar{p})^2} \mu^{2\ve} \int\frac{d^dk}{(2\pi)^d}
	\nonumber\\
	& \quad \times (-g_{T\beta}^\nu)(-g_{T\alpha}^\mu) g_{\mu\nu}^H(p,\bar{p}) g_{\rho\delta}^H(p-k,\bar{p}+k)
\left[ g^{\gamma\delta}(\bar{p}+2k)^\beta - g^{\delta\beta}(2\bar{p}+k)^\gamma + g^{\beta\gamma}(\bar{p}-k)^\delta \right]
	\nonumber\\
	& \quad \times \left[ -g_\gamma^{\;\alpha} (p+k)^\rho + g^{\alpha\rho} (2p-k)_\gamma - g_{\;\gamma}^\rho(p-2k)^\alpha \right]
\frac{1}{\left[(\bar{p}+k)^2+i\Delta^+\right] \left[ (p-k)^2+i\Delta^- \right] \left[ k^2 + i 0 \right] } + h.c.
	\nonumber\\
	& = \frac{\alpha_s C_A}{4\pi} \left[ -\ln^2\frac{i\Delta}{Q^2}  
	+ \frac{13}{12} \left( \frac{1}{\veuv} + \ln\frac{\mu^2}{Q^2} + 2 + i\pi \right) + 
	\frac{5}{2} \left( \frac{1}{\veuv} - \ln\frac{-i\Delta}{\mu^2} + 1 \right) -\frac{17}{18} \right] 
	+ h.c. \, ,
\end{align}

\begin{align}
V^{(\ref{fig:hard}b)} & = 
\frac{-i 2\pi\as C_A}{(d-2)(p\bar{p})^2} \mu^{2\ve} \int\frac{d^dk}{(2\pi)^d}
	\nonumber\\
	& \quad \times (-g_{T\beta}^\nu)(-g_{T}^{\alpha\mu}) g_{\mu\nu}^H(p,\bar{p}) 
		\left[ g^{\beta\rho} (\bar{p}-p+k)^\gamma + g^{\rho\gamma}(p-2k)^\beta - g^{\gamma\beta} (\bar{p}-k)^\rho \right]
	\nonumber\\
	& \quad \times \left[ -g_{\rho\alpha} (2p-k)_\gamma + g_{\gamma\alpha}(p+k)_\rho + g_{\gamma\rho}(p-2k)_\alpha \right]
	\frac{1}{\left[ (p-k)^2+i\Delta^-\right] \left[ k^2 + i 0 \right]} + h.c.
	\nonumber\\
	& = - 
	\frac{\alpha_s C_A}{4\pi} \frac{3}{2} \left[ \frac{1}{\veuv} - \ln\frac{-i\Delta^-}{\mu^2} + 1 \right] + h.c. \, ,
\end{align}

\begin{align}
V^{(\ref{fig:hard}c)} = 0 \,,
\end{align}

\begin{align}
V^{(\ref{fig:hard}d)} & =
\frac{i2\pi\as C_A}{(d-2)(p\bar{p})^2} \mu^{2\ve} \int\frac{d^dk}{(2\pi)^d}
	\nonumber\\
	& \quad\times (-g_{T\beta}^\nu)(-g_{T}^{\mu\alpha}) g_{\mu\nu}^H(p,\bar{p}) g_{\rho\delta}^H(p-k,\bar{p}+k)
	\left[ g_\alpha^{\;\beta} g^{\delta\rho} - g^{\beta\rho}g_\alpha^{\;\delta} 
		- g^{\beta\delta}g_{\alpha}^{\;\rho} + g^{\;\beta}_\alpha g^{\delta\rho} \right]
	\nonumber\\
	&\quad\times \frac{1}{\left[(\bar{p}+k)^2+i\Delta^+\right]\left[(p-k)^2+i\Delta^-\right]} + h.c.
	\nonumber\\
	& =  
	\frac{\alpha_sC_A}{2\pi} \left[ - \frac{13}{12} \left(\frac{1}{\veuv} + \ln\frac{\mu^2}{Q^2} + 2 + i\pi \right) + \frac{17}{18} \right] + h.c.  \,.
\end{align}

Now, adding the contributions in full QCD (with $n_f=5$ flavors) we get:
\begin{align}\label{QCD:full}
 V_{QCD} &=
 2V^{(\ref{fig:hard}a)} + 2V^{(\ref{fig:hard}a)} + V^{(\ref{fig:hard}c)} 
 + V^{(\ref{fig:hard}d)} = 
\frac{\alpha_s C_A}{2\pi} \left[ \frac{2}{\veuv} - 2\ln\frac{\Delta}{\mu^2} 
 - 2\ln^2\frac{\Delta}{Q^2} +\frac{\pi^2}{2} + 2 \right]
 \,.
\end{align}
We have twice the contribution of $V^{(\ref{fig:hard}a)}$ because of the crossed diagram, and twice the contribution of $V^{(\ref{fig:hard}b)}$ because of the two possible diagrams.
Notice that we have not included the calculation of the WFR, because its results will be the same in QCD and in the effective theory and thus will not contribute to the extraction of the hard coefficient.

From Appendix~\ref{app:tmdpdfnlo} we already have the virtual part of the collinear matrix element in SCET (we do not include the WFR and omitting the prefactor $\d(1-x_A)\d^{(2)}(\vecb k_{n\perp})$):
\begin{align}
J_{n} =
\frac{\as C_A}{2\pi} \bigg[
\frac{1}{\veuv} + \ln\frac{\m^2}{\D} + \frac{2}{\veuv}\ln\frac{\D}{Q^2}  
- \ln^2\frac{\D}{Q^2} + 2\ln\frac{\D}{Q^2}\ln\frac{\m^2}{\D} + 1 - \frac{7\pi^2}{12}
\bigg]
\,,
\end{align}
where we have set $\zeta_{A}=Q^2$.
Similarly, for the anti-collinear matrix element in SCET we have (omitting the prefactor $\d(1-x_B)\d^{(2)}(\vecb k_{\bn\perp})$):
\begin{align}
J_{\bn} =
\frac{\as C_A}{2\pi} \bigg[
\frac{1}{\veuv} + \ln\frac{\m^2}{\D} + \frac{2}{\veuv}\ln\frac{\D}{Q^2}  
- \ln^2\frac{\D}{Q^2} + 2\ln\frac{\D}{Q^2}\ln\frac{\m^2}{\D} + 1 - \frac{7\pi^2}{12}
\bigg]
\,,
\end{align}
where we have set $\zeta_{B}=Q^2$.
From Appendix~\ref{app:tmdpdfnlo} we get the virtual part of the soft function in SCET (omitting the prefactor $\d^{(2)}(\vecb k_{s\perp})$):
\begin{align}
S =
\frac{\as C_A}{2\pi} \bigg[
-\frac{2}{\veuv^2} + \frac{4}{\veuv}\ln\frac{\D}{Q^2}
- \frac{2}{\veuv}\ln\frac{\m^2}{Q^2}
- \ln^2\frac{Q^2}{\m^2}
+ 4\ln\frac{\D}{Q^2}\ln\frac{\m^2}{\D}
- \frac{\pi^2}{2}
\bigg]
\,.
\end{align}

Finally, using \eq{eq:facttheoremjjs} and properly including all the deltas in the corresponding prefactors, we obtain the hard coefficient by subtracting to the (renormalized) virtual contribution in full QCD the (renormalized) virtual contributions of the collinear, anti-collinear and soft matrix elements in SCET:
\begin{align}
H(Q^2,\m) &=
1+\big[V_{QCD} - J_n - J_\bn + S\big]_{renormalized}
\nn\\
&=
1+\frac{\as C_A}{2\pi} \bigg[
- \ln^2\frac{Q^2}{\m^2} + \frac{7\pi^2}{6}
\bigg]
\,,
\end{align}
which coincides with the result given in \eq{eq:hardcoeff}.
Notice that we have added the soft function instead of subtracting it, in order to compensate for the double counting of the soft region between the naive (anti)collinear and soft matrix elements.

\section{Anomalous Dimensions}
\label{app:ads}

The anomalous dimension of the top quark Wilson coefficient is given solely by the QCD $\b$-function,
\begin{align}
\g^t(\as(\m)) &=
\frac{d\ln C_t(m_t^2,\mu)}{d\ln\m} =
\as^2 \frac{d}{d\as} \frac{\b(\as(\m))}{\as(\m)}
\,.
\end{align}
Thus we can write the evolution of the coefficient as
\begin{align}
C_t(m_t^2,\mu) &=
\frac{\b(\as(\m))/\as^2(\m)}{\b(\as(\m_0))/\as^2(\m_0)}
C_t(m_t^2,\m_0)
\,.
\end{align}
The coefficient $C_t$ is known up to NNNLO~\cite{Schroder:2005hy,Chetyrkin:2005ia}. 
At NNLO it is~\cite{Kramer:1996iq,Chetyrkin:1997iv}
\begin{align}
C_t(m_t^2,\mu) &= 
1 + \frac{\as(\m)}{4\pi}\,(5C_A-3C_F) 
\nn\\
&
+ \left( \frac{\as(\m)}{4\pi} \right)^2
\bigg[ \frac{27}{2}\,C_F^2 
+ \left( 11\ln\frac{m_t^2}{\mu^2} - \frac{100}{3} \right) C_F C_A 
- \left( 7\ln\frac{m_t^2}{\mu^2} - \frac{1063}{36} \right) C_A^2 
\nn\\
&
- \frac{4}{3}\,C_F T_F - \frac{5}{6}\,C_A T_F 
- \left( 8\ln\frac{m_t^2}{\mu^2} + 5 \right) C_F T_F n_f 
- \frac{47}{9}\,C_A T_F n_f \bigg] 
\,.
\end{align}

The anomalous dimension of the hard part is given by
\begin{align}
\frac{d}{d\ln\mu}\,C_H(-m_H^2,\m) &=
\le[\G^A_{\rm cusp}(\as)\,\ln\frac{-m_H^2}{\m^2} + \g^g(\as) \ri] 
C_H(-m_H^2,\m)
\,
\end{align}
and thus the evolution of the hard coefficient $H=|C_H|^2$ is driven by
\begin{align}
\g_H &=
2\G^A_{\rm cusp}(\as)\,\ln\frac{m_H^2}{\m^2} + 2\g^g(\as)
\,.
\end{align}
The two-loop expression for the Wilson coefficient $C$ can be extracted from the results of \cite{Harlander:2000mg}. 
Writing its perturbative expansion as
\begin{align}
C_H(-m_H^2,\mu) &= 1 + \sum_{n=1}^\infty\,C_n(L)
\left( \frac{\as(\m)}{4\pi} \right)^n
\,,
\end{align}
where $L=\ln[(-m_H^2)/\mu^2]$, the one- and two-loop coefficients are
\begin{align}
C_1(L) &= 
C_A \left( -L^2 + \frac{\pi^2}{6} \right) 
\,,\nn\\\
C_2(L) &= 
C_A^2 \left[ \frac{L^4}{2} + \frac{11}{9}\,L^3
+ \left( -\frac{67}{9} + \frac{\pi^2}{6} \right) L^2 
+ \left( \frac{80}{27} - \frac{11\pi^2}{9} - 2\zeta_3 \right) L
\ri. \nn\\
&
\left.
+ \frac{5105}{162} + \frac{67\pi^2}{36}
+ \frac{\pi^4}{72} - \frac{143}{9}\,\zeta_3 \right]
+ C_F T_F n_f \left( 4L - \frac{67}{3} + 16\zeta_3 \right) 
\nn\\
&
+ C_A T_F n_f \left[ -\frac{4}{9}\,L^3
+ \frac{20}{9}\,L^2 
+ \left( \frac{104}{27} + \frac{4\pi^2}{9} \right) L 
- \frac{1832}{81} - \frac{5\pi^2}{9} - \frac{92}{9}\,\zeta_3 
\right]
\,.
\end{align}
The three-loop result can be extracted from \cite{Baikov:2009bg,Lee:2010cga,Gehrmann:2010tu}.

Below we give the expressions for the anomalous dimensions and the QCD $\beta$-function in the $\overline{{\rm MS}}$ renormalization scheme.
We use the following expansions:
\begin{align}
\G^A_{\rm cusp}(\as) =
\sum_{n=1}^{\infty} \G_{n-1}^A \le( \frac{\as}{4\pi}\ri)^n
\,,
\quad
\g^g(\as) &= \sum_{n=1}^{\infty} \g^g_{n-1} \le( \frac{\as}{4\pi}\ri)^n
\,,
\quad
\g^{nc}(\as) = \sum_{n=1}^{\infty} \g^{nc}_{n-1} \le( \frac{\as}{4\pi}\ri)^n
\,,
\nn\\
\b(\as) &= -2\as \sum_{n=1}^\infty \b_{n-1} \left( \frac{\as}{4\pi} \right)^n
\,.
\end{align}

The cusp anomalous dimension in the adjoint representation can be obtained by multiplying that in the fundamental representation by $C_A/C_F$ (at least up to three-loop order). 
The first three coefficients are
\begin{align}
\G_0^A &=
4 C_A
\,,
\nn\\
\G_1^A &=
4 C_A \le[ \le( \frac{67}{9} - \frac{\pi^2}{3} \ri) C_A - \frac{20}{9}\,T_F n_f \ri]
\,,
\nn\\
\G_2^A &=
4 C_A \le[ C_A^2 \left( \frac{245}{6} - \frac{134\pi^2}{27}
+ \frac{11\pi^4}{45} + \frac{22}{3}\,\zeta_3 \right)
+ C_A T_F n_f  \left( - \frac{418}{27} + \frac{40\pi^2}{27}
- \frac{56}{3}\,\zeta_3 \right)
\ri.
\nn\\
&\le.
+ C_F T_F n_f \left( - \frac{55}{3} + 16\zeta_3 \right)
- \frac{16}{27}\,T_F^2 n_f^2 \ri]
\,.
\end{align}

The first three coefficients of the anomalous dimension $\g^g$ are~\cite{Idilbi:2005ni,Idilbi:2005er}
\begin{align}
\g^g_0 &= 0 
\,, \nn\\
\g^g_1 &= 
C_A^2 \left( -\frac{160}{27} + \frac{11\pi^2}{9}
+ 4\zeta_3 \right) 
+ C_A T_F n_f \left( -\frac{208}{27} - \frac{4\pi^2}{9} \right) 
- 8 C_F T_F n_f 
\,, \nn\\
\g^g_2 &= 
C_A^3 \left[ \frac{37045}{729} + \frac{6109\pi^2}{243}
- \frac{319\pi^4}{135} 
+ \left( \frac{244}{3} - \frac{40\pi^2}{9} \right) \zeta_3 
- 32\zeta_5 \right] 
\nn\\
&
+ C_A^2 T_F n_f \left( -\frac{167800}{729}
- \frac{2396\pi^2}{243} + \frac{164\pi^4}{135} 
+ \frac{1424}{27}\,\zeta_3 \right)
\nn\\
&
+ C_A C_F T_F n_f \left( \frac{1178}{27}
- \frac{4\pi^2}{3} - \frac{16\pi^4}{45} 
- \frac{608}{9}\,\zeta_3 \right) + 8 C_F^2 T_F n_f
\nn\\  
&
+ C_A T_F^2 n_f^2 \left( \frac{24520}{729}
+ \frac{80\pi^2}{81} - \frac{448}{27}\,\zeta_3 \right) 
+ \frac{176}{9} C_F T_F^2 n_f^2 
\,.
\end{align}

The first three coefficients of $\g^{nc}$ are
\begin{align}
\g_0^{nc} &= 
- 2\b_0 = 
- \frac{22}{3}\,C_A + \frac{8}{3}\,T_F n_f 
\,, 
\nonumber\\
\g_1^{nc} &= 
2C_A^2 \left( -\frac{692}{27} + \frac{11\pi^2}{18}
+ 2\zeta_3 \right) 
+ 2C_A T_F n_f \left( \frac{256}{27} - \frac{2\pi^2}{9} \right)
+ 8 C_F T_F n_f 
\,, 
\nonumber\\
\g_2^{nc} &= 
2C_A^3 \left( - \frac{97186}{729} 
+ \frac{6109\pi^2}{486} - \frac{319\pi^4}{270} 
+ \frac{122}{3}\,\zeta_3 - \frac{20\pi^2}{9}\,\zeta_3 
- 16\zeta_5 \right) 
\nonumber\\
&\mbox{}
+ 2C_A^2 T_F n_f \left( \frac{30715}{729}
- \frac{1198\pi^2}{243} + \frac{82\pi^4}{135} 
+ \frac{712}{27}\,\zeta_3 \right) 
\nonumber\\
&\mbox{}
+ 2C_A C_F T_F n_f \left( \frac{2434}{27} 
- \frac{2\pi^2}{3} - \frac{8\pi^4}{45} 
- \frac{304}{9}\,\zeta_3 \right) 
- 4 C_F^2 T_F n_f 
\nonumber\\
&\mbox{}
+ 2C_A T_F^2 n_f^2 \left( - \frac{538}{729}
+ \frac{40\pi^2}{81} - \frac{224}{27}\,\zeta_3 \right) 
- \frac{88}{9}\,C_F T_F^2 n_f^2 \,.
\end{align}

Finally, the coefficients for the QCD $\beta$-function are
\begin{align}
\b_0 &=
\frac{11}{3}\,C_A - \frac43\,T_F n_f
\,,
\nn\\
\b_1 &=
\frac{34}{3}\,C_A^2 - \frac{20}{3}\,C_A T_F n_f
- 4 C_F T_F n_f
\,,
\nn\\
\b_2 &=
\frac{2857}{54}\,C_A^3 + \left( 2 C_F^2
- \frac{205}{9}\,C_F C_A - \frac{1415}{27}\,C_A^2 \right) T_F n_f
+ \left( \frac{44}{9}\,C_F + \frac{158}{27}\,C_A \right) T_F^2 n_f^2
\,,
\nn\\
\b_3 &=
\frac{149753}{6} + 3564\zeta_3
- \left( \frac{1078361}{162} + \frac{6508}{27}\,\zeta_3 \right) n_f
+ \left( \frac{50065}{162} + \frac{6472}{81}\,\zeta_3 \right) n_f^2
+ \frac{1093}{729}\,n_f^3
\,,
\end{align}
where for $\b_3$ we have used $N_c=3$ and $T_F=\frac{1}{2}$.


\end{document}